\documentclass[twocolumn,letterpaper,amsmath,amssymb,amsfonts,floatfix,aps,superscriptaddress,showkeys,showpacs,nofootinbib]{revtex4-1}
 
\usepackage{hyperref} 
\usepackage{verbatim} 
\usepackage{graphicx}
\usepackage{bm} 
\usepackage{mathtools}
\usepackage{ifpdf} 
\usepackage{dcolumn}  
\usepackage{epsfig} 
\usepackage{color}
\usepackage[usenames,dvipsnames]{xcolor} 
\usepackage{multirow}
\usepackage{float}
\usepackage[makeroom]{cancel} 

\newcommand{\sgn}{\operatorname{sgn}}
\newcommand{\icomplex}{\dot\iota}
\newcommand{\real}{\operatorname{Re}}
\newcommand{\imaginary}{\operatorname{Im}}

\newcommand{\wchi}{\widetilde{\chi}}

\newcommand{\wf}{\widetilde{f}}
\newcommand{\wv}{\widetilde{v}}

\newcommand{\wu}{\widetilde{u}}

\newcommand{\omegan}{\omega_{\!_\perp}} 
\newcommand{\taun}{\tau_{\!_\perp}}
\newcommand{\wchinorm}{\wchi_{\!_\perp}}
\newcommand{\wcalGnorm}{\widetilde {\mathcal G}_{\!_\perp}}
\newcommand{\cnorm}{\mathcal{C}_{\!_\perp}}
\newcommand{\wcnorm}{{\widetilde{\mathcal{C}}}_{\!_\perp}}
\newcommand{\nun}{\nu_{\!_\perp}}
\newcommand{\gamman}{\gamma_{\!_\perp}}
\newcommand{\hn}{h_{\!_\perp}}
\newcommand{\zn}{z_{\!_\perp}}
\newcommand{\an}{a_{\!_\perp}}

\newcommand{\omegas}{\omega_{\!_\parallel}}
\newcommand{\taus}{\tau_{\!_\parallel}}
\newcommand{\wchishear}{\wchi_{\!_\parallel}}
\newcommand{\wcalGshear}{\widetilde {\mathcal G}_{\!_\parallel}}
\newcommand{\cshear}{\mathcal{C}_{\!_\parallel}}
\newcommand{\wcshear}{{\widetilde{\mathcal{C}}}_{\!_\parallel}}
\newcommand{\nus}{\nu_{\!_\parallel}}
\newcommand{\gammas}{\gamma_{\!_\parallel}}
\newcommand{\hs}{h_{\!_\parallel}}
\newcommand{\zs}{z_{\!_\parallel}}
\newcommand{\as}{a_{\!_\parallel}}

\newcommand{\omegasn}{\omega_{\!_{\parallel,\perp}}}

\newcommand{\nusn}{\nu_{\!_{\parallel,\perp}}}
\newcommand{\asn}{a_{\!_{\parallel,\perp}}}

\begin{document}

\title{Hydrodynamic stress correlations in fluid films driven by stochastic surface forcing} 

\author{Masoud \surname{Mohammadi-Arzanagh}} 
\affiliation{School of Physics, Institute for Research in Fundamental Sciences (IPM), P.O. Box 19395-5531, Tehran, Iran
}
\affiliation{Department of Physics, Sharif University of Technology,  P.O. Box 11155-9161, Tehran, Iran}
\author{Saeed \surname{Mahdisoltani}} 
\affiliation{School of Physics, Institute for Research in Fundamental Sciences (IPM), P.O. Box 19395-5531, Tehran, Iran
}
\affiliation{Department of Physics, Sharif University of Technology,  P.O. Box 11155-9161, Tehran, Iran}
\author{Rudolf \surname{Podgornik}}
\affiliation{School of Physical Sciences and Kavli Institute for Theoretical Sciences,
University of Chinese Academy of Sciences, Beijing 100049, China}
\affiliation{CAS Key Laboratory of Soft Matter Physics, Institute of Physics,
Chinese Academy of Sciences, Beijing 100190, China}
\author{Ali \surname{Naji}}\thanks{Corresponding author: a.naji@ipm.ir}
\affiliation{School of Physics, Institute for Research in Fundamental Sciences (IPM), P.O. Box 19395-5531, Tehran, Iran
}

\begin{abstract}
We study hydrodynamic fluctuations in a compressible and viscous fluid film confined between two rigid, no-slip, parallel plates, where one of the plates is kept fixed, while the other one is driven in small-amplitude, translational, displacements around its reference position. This jiggling motion is assumed to be driven by a stochastic, external, surface forcing of zero mean and finite variance. Thus, while the transverse (shear) and longitudinal (compressional) hydrodynamic stresses produced in the film vanish on average on either of the plates, these stresses exhibit fluctuations that can be quantified through their equal-time, two-point, correlation functions. For transverse stresses, we show that the correlation functions of the stresses acting on the same plate (self-correlators) as well as the correlation function of the stresses acting on different plates (cross-correlators) exhibit universal,  decaying, power-law behaviors as functions of the inter-plate separation. At small separations, the exponents are given by $-1$, while at large separations, the exponents are found as $-2$ (self-correlator on the fixed plate), $-4$ (excess self-correlator on the mobile plate) and $-3$ (cross-correlator). For longitudinal stresses, we find much weaker power-law decays  in the large separation regime, with exponents $-3/2$ (excess self-correlator on the mobile plate) and  $-1$ (cross-correlator). The self-correlator  on the fixed plate increases and levels off upon increasing the inter-plate separation, reflecting the non-decaying nature of the longitudinal forces acting on the fixed plate. 
\end{abstract}

\maketitle

\section{Introduction}

Effective interactions between nano-/macromolecular bodies in aqueous solutions can broadly be decomposed into two equally important contributions: The static or equilibrium forces, and the dynamic forces, arising when the system is driven out of equilibrium, such as when the bodies are in relative motion \cite{Israela,Butt}. While the former originate in the disjoining pressure due to direct and/or solvent-mediated surface forces, operating primarily at the nanoscale, the latter depend on dynamic molecular processes in the solvent, and also on (slow) hydrodynamic stresses as the intervening solvent is drained/sheared from the liquid film separating the interacting surfaces. The dynamic forces can thus be relevant over a much wider range of nano-/microscale separations \cite{Brenner,Butt,Israela,Dhont}. 

The extended Derjaguin-Landau-Verwey-Overbeek theory of colloidal stability identifies three types of static surface interactions \cite{Israela,Butt}: The electrostatic interactions depending on the specific nature of mobile and fixed molecular charges \cite{Elec}, the ubiquitous van-der-Waals (vdW) interactions, depending on the dielectric response of molecular materials \cite{Woods}, and solvent-mediated interactions stemming from the hydrophobic and/or hydration forces between solvent-exposed surfaces \cite{Hydration}.  
The dynamic forces are, on the other hand, much more difficult to classify unequivocally. Some of the equilibrium forces, as is the case with, for instance, the vdW interaction itself, can display an inherent dynamic component \cite{Kardar,Mkrtchian}. Others may exhibit no equilibrium counterpart as are the hydrodynamic interactions, having significant impact on dynamic properties (e.g., spatiotemporal correlations) of colloids in bulk \cite{Brenner,Butt,Israela,Dhont} or strongly confined fluids \cite{Keyser,Diamant}. 

Recent advances in surface-force techniques, such as surface forces apparatus (SFA) and atomic force microscope (AFM) \cite{Israela,Butt}, have enabled high-precision determination of both static and dynamic forces acting between contact surfaces across an intervening layer of simple or complex fluid (for recent reviews of relating techniques and applications, see  Refs.  \cite{Butt-rev,Israelachvili_ROPP,Korea,Haviland,Ellis3,Neto,Bocquet2010,Wang2017}). Dynamic SFA usually  incorporates two apposed, molecularly smooth, flattened or curved surfaces of relatively large radii of curvature,  with one of the surfaces driven in controlled three-dimensional (linear/oscillating) motion \cite{Israelachvili_ROPP}. This allows for measuring various (generally frequency-dependent) rheological properties of thin fluid films and gives direct access to shear/compressional forces exerted on the bounding surfaces by the intervening fluid over a wide range of surface separations and velocities/frequencies of the imposed surface motion \cite{Granick1991b,Klein1998,Klein2007,CottinBizonne2008,Bureau2010,Leroy2012,Steinberger2008}. Dynamic AFM 
has, on the other hand, emerged as an important tool for probing the local response of hard or soft material interfaces in liquid media \cite{Butt-rev,Haviland,Wang2017}. In colloidal-probe AFM  \cite{Butt-rev}, a relatively large, cantilever-mounted colloidal particle oscillates in proximity to an interface, with the power spectrum of the oscillations providing information on the hydrodynamic/viscoelastic properties of the surrounding liquid and the probe-surface interactions.  
The effects of oscillatory external forcing as well as thermal noise in dynamic AFM, involving  colloidal probes or flat microlevers, have been analyzed on various levels of approximation   \cite{Sader3,Alcaraz,Benmouna,Butt-rev,Clarke,Siria2009}. In very recent works,  Maali et al. have first analyzed the hydrodynamics of a vibrating microsphere \cite{Maali1} and then generalized the methodology to a thermally driven vibrating sphere yielding a thermal-noise AFM probe, where the sub-nanometer 
thermal motion of the sphere, coupled to a spring and dashpot mathematical model, reveals an (elasto)hydrodynamic coupling between the sphere and a vicinal, hard (mica) or soft (air-bubble), substrate in water \cite{Maali2}. 

While shear/drainage thin-film flows, caused by small-amplitude oscillatory or stochastic surface forcing \cite{Israelachvili_ROPP,Korea,Butt-rev,Haviland,CottinBizonne2008,Leroy2012,Steinberger2008,Wang2017,Maali2,Maali1,Butt-rev,Alcaraz,Benmouna,Clarke,Sader3,Siria2009}, have been a common motif in the SFA/AFM contexts, other techniques for generating such flow patterns have been developed based on quartz crystal resonators (QCRs) to probe near-surface fluid properties  \cite{Butt}, such as boundary slippage effects \cite{Ellis3,Neto,Bocquet2010}. QCRs are  used (also in combination with the dynamic SFA \cite{Berg2003}) as vibrating fluid  substrates, driven at their resonance frequency to produce unsteady thin-film flows at  high frequencies and shear rates (see the review in Ref. \cite{Neto}). 

In the aforementioned contexts, it is important to analyze first the well-defined limits and only then proceed to more advanced models to account for the various couplings and feedbacks \cite{Grier}. It is also important to realize 
that forced motion is in general incompatible with the assumptions of weak acceleration, requiring one to account for finite compressibility effects, which makes the underlying hydrodynamic problem more difficult to tackle \cite{Netz}. 

Motivated by these advances, we formulate a general framework for surface-driven hydrodynamic interactions across a compressible and viscous fluid film, mechanically driven,  in transverse/longitudinal directions, at one of its two rigid boundaries using an  arbitrary external  forcing. We focus on a fluid film with plane-parallel bounding surfaces, which is more yielding towards systematic calculations. Our primary interest is in  the fluctuational behavior of shear/compressional hydrodynamic stresses generated across the film and on the boundaries, when the surface forcing is stochastic and of arbitrary (thermal or non-thermal) spectral density. The concrete example of temporally uncorrelated (white-noise) forcing will be analyzed numerically in detail. 
We show that the compressible, viscous hydrodynamic coupling between the mobile (forced) and the fixed bounding surfaces of the film leads to a complicated dependence of the same-plate and cross-plate stress correlation functions on the inter-surface separation (film thickness). This includes decaying power-laws with universal exponents and {\em non-decaying} stress variances on the fixed plate due to the acoustic resonances originating in the compressional modes. 

This problem in some sense represents an inverse one with respect to the recently analyzed case of thermal  fluctuating hydrodynamics \cite{LL} between two rigid plates, geared towards elucidating the possible role of the {\em hydrodynamic Casimir-like effects} (in analogy with other examples of non-equilibrium fluctuation-induced forces involving fluctuating classical fields \cite{Jones,Chan, Monahan, Bartolo,Dean,Ajay1, Antezza, Kruger, Kirkpatrick13, Kirkpatrick14, Kafri-Kardar}). The hydrodynamic Casimir-like phenomenon was in fact shown to exist only in its indirect ``secondary" form \cite{Monahan}: While the average stress on either of the bounding surfaces is zero, its fluctuations indicate long-range correlations, 
as a result of near-equilibrium thermal fluctuations in the confined fluid film. The forced fluctuations of one of the bounding plates of a confined fluid layer analyzed below are rather different from the hydrodynamic Casimir-effect phenomenology, and are more closely related to the so-called {\em Bjerknes interactions} in driven acoustic resonators \cite{Bruus2,Leighton}. Contrary to the Bjerknes interactions though, the forcing  in our case is not due to a volume-distributed external acoustic field, but is rather exerted on one (or both) of the rigid bounding plates.

We introduce our framework in Section \ref{sec:formalism} and calculate the relevant hydrodynamic response/correlations for an arbitrary stochastic surface forcing
in Section  \ref{sec:response}. The numerical results are given for the  special case of white-noise  forcing in Sections \ref{sec:shear_corr} and \ref{sec:norm_corr}, followed by the conclusions in Section \ref{sec:conclusion}.

\section{Model and Formalism}
\label{sec:formalism}

\subsection{Model geometry and physical description}
\label{sec:model}
 
Let us consider a classical, compressible, viscous fluid film confined between two rigid, parallel plates of infinite extent in the $x-y$ coordinate plane at vertical locations $z=0$ and $z=h>0$; see Fig. \ref{fig:schematic}. The upper plate is kept at rest, while an 
{external surface force {\em per unit area} $\mathbf{f}=\mathbf{f}(t)$, drives the lower plate in arbitrary, translational, rigid-body movements around its reference plane at $z=0$. As a consequence the mobile plate exhibits a time-dependent {\em surface velocity}, $\mathbf{u}(t)$, which is to be determined consistently and concurrently with the fluid velocity, density and pressure fields within the film, i.e., $\mathbf{v}=\mathbf{v}(\mathbf{r}; t)$, $\rho=\rho(\mathbf{r}; t)$, and $p=p(\mathbf{r}; t)$, respectively.

In its general aspects, the present model is used to capture the elemental features of  typical surface-driven flows in standard surface-force experiments \cite{Israelachvili_ROPP,Korea,Butt-rev,Haviland,CottinBizonne2008,Leroy2012,Steinberger2008,Wang2017,Maali2,Maali1,Butt-rev,Alcaraz,Benmouna,Clarke,Sader3,Siria2009,Ellis3,Neto,Bocquet2010,Granick1991b,Klein1998,Klein2007,Bureau2010,Berg2003,Grier}, but it is also designed as a first-step model to facilitate an unequivocal elucidation of the basic physics of the problem using direct analytical calculations. It is nevertheless useful to detail the simplifying assumptions involved. 

First, we note that while flattened/planar contact surfaces, as in our model, have been used in dynamic SFA/QCRs (see, e.g., Refs. \cite{Butt,Israela,Granick1991b,Klein1998,Berg2003,Israelachvili_ROPP,Bureau2010}), and also in dynamic AFM with wide flat microlevers \cite{Siria2009}, experiments often rely on cross-cylindrical or sphere-plane geometries. 
The radii of curvature in these applications are however very large (around a few mm/cm in SFA and tens of $\mu$m in AFM) as compared to the film thickness (varied in the sub-nm to $\mu$m range); see, e.g., Refs. \cite{Israelachvili_ROPP,Granick1991b,Klein1998,Klein2007,CottinBizonne2008,Steinberger2008,Bureau2010,Korea,Butt-rev,Haviland,Wang2017,Alcaraz,Benmouna,Maali1,Maali2,Leroy2012}. For such weakly curved surfaces, the Derjaguin approximation \cite{Israela,Butt} can be used to predict the interaction forces between curved boundaries based merely on the results obtained in the plane-parallel geometry, or vice versa. 

Secondly, the amplitude of displacements (especially in $z$ direction) of the mobile plate is assumed here to be much smaller than the film thickness or, equivalently, $|\mathbf{u}(t)|$ is taken be sufficiently small \cite{Note_small_parameter}. This is in fact the typical situation also for the perpendicular (compressional), oscillatory or noise-driven, surface motions  utilized in dynamic SFA/AFM experiments \cite{Israelachvili_ROPP,Korea,Butt-rev,Haviland,CottinBizonne2008,Leroy2012,Steinberger2008,Wang2017,Maali2,Maali1,Alcaraz,Benmouna,Clarke,Sader3,Siria2009}. It enables one to assume that the inter-plate separation is fixed on the leading order and also  allows for a linearized treatment of the full Navier-Stokes equations \cite{Alcaraz,Benmouna,Clarke,Sader3}  by setting $\mathbf{v} = \mathbf{v}^{(1)}$, $p = p_0+p^{(1)}$ and $\rho = \rho_0+\rho^{(1)}$, where the superscript $(1)$ denotes the first-order  fluctuations around the rest values $\mathbf{v}=\mathbf{0}$, $p=p_0$ and $\rho =\rho_0$. 

Thirdly, we neglect possible boundary slippage effects \cite{Ellis3,Neto,Bocquet2010,CottinBizonne2008,Maali1,Maali2,Siria2009,Steinberger2008} by taking no-slip boundaries with $ \mathbf{v}(x, y, z=0; t) = \mathbf{u}(t)$ and $ \mathbf{v}(x, y, z=h; t) = \mathbf{0}$. 

Finally, we ignore local temperature variations and heat transfer processes in the film \cite{Note_T} (to be considered elsewhere \cite{Chris2017}), the (nonlinear) viscous dissipation \cite{Klein2007}, and  the relaxation effects that can formally be accounted for by taking frequency-dependent viscosities \cite{LL}. 
  
\begin{figure}[t!]
\begin{center}
\includegraphics[width=7cm]{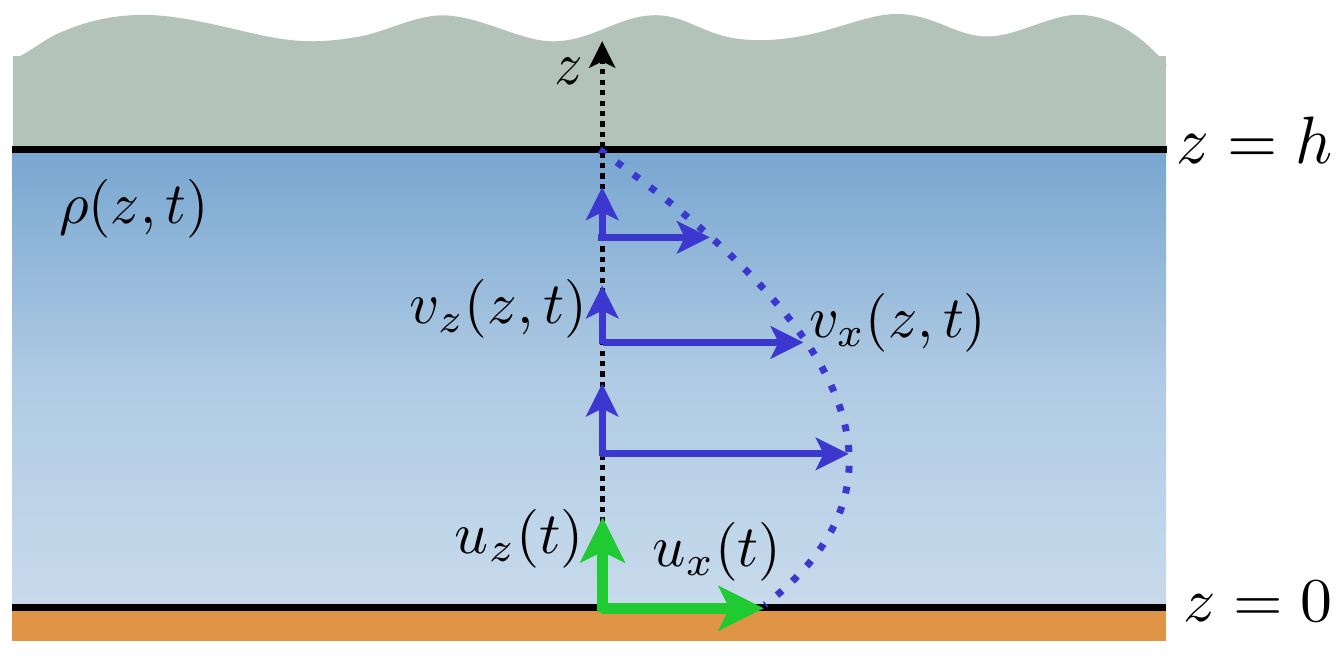}
\caption{Sideview of a plane-parallel film of a compressible, viscous fluid, driven at its lower boundary ($z=0$) with an external forcing, giving the uniform surface velocity $\mathbf{u}(t)$. }
\label{fig:schematic}
\end{center}
\end{figure}

\subsection{Linearized fluid-film hydrodynamics}
\label{sec:linhydro}

The surface-driven flow problem described above is governed by the following set of equations to the first order in field fluctuations \cite{LL}
\begin{align}
&\!\!\!\!\eta \nabla^2 \mathbf{v}^{(1)}\! +\! \left( \frac{\eta}{3} + \zeta\right)\!
\nabla\left(\!\nabla\cdot \mathbf{v}^{(1)}\!\right)\! -\! \nabla 
	p^{(1)}\!-\! \rho_0\partial_t
	\mathbf{v}^{(1)} \!  = 0,  
\label{eq:lns1}\\
&\quad\,\, \partial_t\rho^{(1)}+ 	     			    
	\rho_0\nabla \cdot \mathbf{v}^{(1)} {} = 0,
\qquad \,\, p^{(1)} = c_0^2\rho^{(1)},
\label{eq:prho}
\end{align}
where $ c_0$ 
is the isothermal speed of sound \cite{Note_typo}. These equations are supplemented by the no-slip boundary conditions $\mathbf{v}^{(1)}  (x,y, z=0; t) = \mathbf{u}(t)$, $\mathbf{v}^{(1)} (x, y, z=h; t) = 0$.  The first-order hydrodynamic stress tensor is
\begin{equation}
\label{eq:sigma}
\sigma_{jk}^{(1)} \!=  \eta \big[\nabla_j v_k^{(1)} + \nabla_k v_j^{(1)}\big] - \delta_{jk}\!\left[\left(\frac{2\eta}{3} - \zeta\right) \!\nabla_l v_l^{(1)} +c_0^2 \rho^{(1)}\right], 
\end{equation}
with $j, k, l=x, y, z$ denoting the Cartesian components.

We drop the superscript $(1)$ for notational simplicity; thus, $\mathbf{v}(\mathbf{r}; t)$, $\rho(\mathbf{r}; t)$, and $p(\mathbf{r}; t)$ hereafter denote {\em only} the first-order field fluctuations around the given stationary values.  Due to the one-dimensional nature of the flow, we  drop the variables $x$ and $y$, and use the notation $\mathbf{v} = (\mathbf{v_{\!_\parallel}}, v_z)$, $\mathbf{v_{\!_\parallel}} = (v_x, v_y)$ to simplify Eqs. (\ref{eq:lns1}) and (\ref{eq:prho}) as 
\begin{align}
&\rho_0 \partial_t \mathbf{v_{\!_\parallel}} = \eta \partial_z^2 \mathbf{v_{\!_\parallel}}, 
	\label{eq:linvx}\\
	&\rho_0 \partial_t v_z = -c_0^2 \partial_z \rho + 			
			\left(\frac{4 \eta}{3} + \zeta\right) \partial_z^2 v_z,					    
	\label{eq:linvz}\\
		&\partial_t \rho + \rho_0 \partial_z v_z=0.
	\label{eq:linrho}
\end{align}
The different components of the velocity field are thus decoupled, as can be seen by combining Eqs. (\ref{eq:linvz}) and (\ref{eq:linrho}) to obtain the standard, attenuated wave equation 
\begin{equation} 
\label{eq:zwave}
\partial^2_t v_z = c_0^2 \partial_z^2 v_z + \nun {\partial_z^2 \partial_t v_z}. 
\end{equation}
Due to the symmetry upon interchanging $v_x$ and $ v_y$, we restrict our  discussions of $\mathbf{v_{\!_\parallel}}$ only to its $v_x$ component. The relevant non-zero components of the stress tensor include only the transverse and longitudinal ones
\begin{align}
\label{eq:sigmax_0}
\sigma_{xz}(z;t)&= \rho_0\nus \partial_z v_x\\
\label{eq:sigmaz_0}
\sigma_{zz}(z;t)&= \rho_0 \nun \partial_z v_z - c_0^2 \rho (z;t),
\end{align}
where $\nus = \eta/\rho_0$ and $\nun = \left(4\eta/3+\zeta\right)/\rho_0$ are the corresponding  transverse (shear) and longitudinal  (compressional) kinematic viscosities, respectively. 

Since our formulation is linear, the solutions reported below can linearly be superposed to study the case with {\em both} lower and upper plates undergoing small-amplitude displacements. This is a straightforward generalization, which we shall not discuss any further.

\subsection{Equation of motion for the mobile surface}
\label{subsec:EOM}

Since the lower (mobile) plate is driven by the external force per unit area, $\mathbf{f}=\mathbf{f}(t)$, causing it to move with the velocity $\mathbf{u}=\mathbf{u}(t)$, we can write Newton's second law for its motion in the frequency (Fourier) domain as
\begin{equation}
 \label{eq:eom}
- {\icomplex} m \omega \wu_j(\omega) = \wf_j(\omega) - \widetilde{\sigma}_{jz}\big(z=0; \omega\big) n_z,\quad j=x, z, 
\end{equation}
where $m$ is the plate mass {\em per unit area}, and $\wf_j$ and $\wu_j$ are the frequency-domain components of $\mathbf{f}$ and $\mathbf{u}$, respectively. 
Here, $n_z=-1$ is the $z$ component of the unit vector along the inward normal to the mobile plate, making $-\widetilde{\sigma}_{jz}\left(z=0; \omega\right) n_z$ the force component per unit area acting on the plate due to hydrodynamic stresses \cite{LL}. 

We solve Eqs. (\ref{eq:linvz}) and (\ref{eq:linrho}) with the required boundary conditions to find the solutions for the fluid velocity/density  fields in terms of the surface velocity $\mathbf{u}(t)$, which can itself be determined as a function of $\mathbf{f}(t)$ by inserting those solutions into the stress term in Eq. (\ref{eq:eom}). This gives the the desired final forms of the fluid velocity/density fields as functions of the external forcing $\mathbf{f}(t)$. 
 
Our formulation can be implemented with any surface forcing model of either deterministic or stochastic origins. Stochastic forcing is particularly relevant to the thermal-noise AFM probe \cite{Maali2}, resulting, e.g., from Brownian fluctuations in the setup or ambient fluid. We shall adopt a Gaussian-distributed stochastic forcing with mean $\langle \wf_j(\omega) \rangle = 0$ and two-point correlation function $\langle \wf_j(\omega) \wf_k(\omega') \rangle = 4\pi {\widetilde {\mathcal G}_j}(\omega) \delta_{jk}\delta(\omega+\omega')$, 
where ${\widetilde {\mathcal G}_j}(\omega)$ are the real-valued and positive forcing spectral densities. In our numerical analysis later, we shall adopt the {\em white-noise Ansatz} with ${\widetilde {\mathcal G}_j}(\omega) = {\mathcal G}_j$  (of dimension $[{\textrm{pressure}}]^2\cdot [{\textrm{time}}]$)  taken as  constants. 

Note that the {\em zero mean} taken for the surface forcing implies that the hydrodynamic  stresses acting on the plates due to the fluctuations in the film will be zero on average, 
$\langle \sigma_{xz}(z;t)\rangle = \langle\sigma_{zz}(z;t) \rangle = 0$,  
where the brackets $\langle\cdots\rangle$ denote ensemble averaging over various realizations of the external forcing. Yet, the variance and correlation functions (or correlators) of instantaneous stresses can be finite, characterizing the measurable fluctuation-induced forces mediated by hydrodynamic correlations between the plates on the leading order. Although, these forces resemble the secondary hydrodynamic Casimir-like forces arising from near-equilibrium thermal fluctuations \cite{Monahan}, it is important to note that the stochastic forcing here can generally be non-thermal, in which case it can produce far-from-equilibrium stress fluctuations and correlations in the fluid film. To evaluate these  correlations, we first give the solutions to the velocity and density fields in terms of the stress response functions of the fluid film.

\section{Response to surface forcing}
\label{sec:response}

\subsection{Velocity and density fields} 
\label{subsec:vel_dens_fields}

The velocity and density field fluctuations can be obtained by transforming Eqs.~\eqref{eq:linvx} and \eqref{eq:zwave} 
to the frequency domain. The governing equations for transverse and longitudinal modes read, respectively,  as
\begin{align} 
\label{eq:vxeq}
&\partial_z^2 \wv_x (z;\omega) = -\alpha^2(\omega)\,\wv_x (z;\omega), \\
&\partial_z^2 \wv_z (z;\omega) = -\kappa^2(\omega)\,\wv_z(z;\omega),
\label{eq:vzeq}
\end{align}
where  $\alpha^2(\omega) = {\icomplex}{\omega}/{\nus}$ and $\kappa^2(\omega) = {\omega^2}/{\left(c_0^2 - {\icomplex} \omega \nun\right)}$.  
Here, $\alpha$ and $\kappa$ give the frequency-dependent (screening) length-scales  associated with the shear and compressional modes, respectively. There will be two equivalent sets of solutions for  $\alpha$ and $\kappa$, fulfilling the relations $\alpha^\ast(\omega)=\pm\alpha(-\omega)$, $\kappa^\ast(\omega)=\pm\kappa(-\omega)$. One can conveniently choose the solutions satisfying these relations with plus signs, giving  real (R) and imaginary (I) parts
\begin{align} 
\label{eq:alphaRI}
&\alpha_{\mathrm{R}}(\omega) =  \alpha_{\mathrm{I}}(\omega)\sgn(\omega)  = \pm \sqrt{\frac{|\omega|}{2 \nus}},\\
&\kappa_{\mathrm{R}}(\omega) = \pm\frac{|\omega|}{\sqrt{2}} \sqrt{\frac{\sqrt{c_0^4 + \omega^2 \nun^2}+c_0^2}{c_0^4 + \omega^2 \nun^2}},
\label{eq:kappaR}\\
&\kappa_{\mathrm{I}}(\omega) = \pm\frac{\omega}{\sqrt{2}} \sqrt{\frac{\sqrt{c_0^4 + \omega^2 \nun^2}-c_0^2}{c_0^4 + \omega^2 \nun^2}},
\label{eq:kappaI}
\end{align}
where $\sgn(\cdot)$ is the sign function.  
Now, solving Eqs.~(\ref{eq:vxeq}) and~(\ref{eq:vzeq}) with the required boundary conditions gives
\begin{align}
v_x (z;t) &= \int \frac{\mathrm{d}\omega}{2\pi}\,e^{-{\icomplex}\omega t}\,\frac{\sin\left[\alpha(\omega) (h-z)\right]}{\sin\left[\alpha(\omega) h \right]}\,\wu_x(\omega),\label{eq:vxsol}\\
v_z (z;t) &= \int \frac{\mathrm{d}\omega}{2\pi}\,e^{-{\icomplex}\omega t}\,\frac{\sin\left[\kappa(\omega) (h-z)\right]}{\sin\left[\kappa(\omega) h\right]}\,\wu_z(\omega). \label{eq:vzsol}
\end{align}
The density field fluctuations can be obtained 
using Eqs.~(\ref{eq:linrho}) 
and Eq. \eqref{eq:vzsol}, yielding 
\begin{align}
\label{eq:rhointegral}
\rho(z;t) &= \rho_0\!\int \frac{\mathrm{d}\omega}{2\pi}\,e^{-{\icomplex}\omega t} \left[\frac{{\icomplex}\kappa(\omega)}{\omega}\right]  \frac{\cos\left[\kappa(\omega) (h-z)\right]}{\sin\left[\kappa(\omega) h\right]}\,\wu_z(\omega).
\end{align}

It should be noted that the homogeneous parts of the solutions  (in the time domain) are discarded  as they depend on initial conditions, being irrelevant in the long-time stationary state to be studied here. 

\subsection{Response and correlation functions} 
\label{subsec:response_functions}

Plugging Eqs.  \eqref{eq:vxsol}-\eqref{eq:rhointegral} into Eqs.  \eqref{eq:sigmax_0} and \eqref{eq:sigmaz_0}, we can write the two components of the surface stress tensor, $\widetilde{\sigma}_{jz}(z=0; \omega)$, in terms of the surface velocity components $\wu_j(\omega)$ for $j=x, z$. Inserting the results into Eq. (\ref{eq:eom}), we find $\wu_j(\omega)$  in terms of the surface forcing components, $\wf_j(\omega)$ and, then,  using Eqs. \eqref{eq:vxsol} and \eqref{eq:vzsol},  find
the velocity field components $v_j (z;t)$  in terms of $\wf_j(\omega)$ as  
\begin{align}
v_j (z;t) &= \int \frac{\mathrm{d}\omega}{2\pi}\,e^{-{\icomplex}\omega t}\, {\widetilde R}_j(z; \omega)\wf_j(\omega), 
\label{eq:vxsol1}
\end{align}
where the {\em velocity response functions} are given by
\begin{widetext}
\begin{align}
{\widetilde R}_x(z; \omega) &= \frac{\sin{[\alpha(\omega)(h-z)]}}{\rho_0\nus \alpha(\omega) \left( \cos{[\alpha(\omega)h]} - \frac{m}{\rho_0}\alpha(\omega) \sin{[\alpha(\omega)h]}\right)}, 
\label{eq:vxsol2}\\
{\widetilde R}_z(z; \omega) &= \frac{\sin{[\kappa(\omega)(h-z)]}}{\rho_0\nun \left( 1 + {\icomplex}\frac{c_0^2}{\nun\omega}\right)\kappa(\omega) \left( \cos{[\kappa(\omega)h]} - \frac{m}{\rho_0}\kappa(\omega) \sin{[\kappa(\omega)h]}\right)}. 
\label{eq:vzsol2}
\end{align}
\end{widetext}
These play the role of the Oseen tensor components in the considered geometry.   We can now express the stress tensor components   in terms of the external forcing as 
\begin{align} 
\label{eq:sigmaxf1}
\sigma_{jz}(z;t)&=\int\frac{\mathrm{d}\omega}{2\pi}\,e^{-{\icomplex}\omega t} \wchi_{jz}(z; \omega)\wf_{j}(\omega), 
\end{align} 
where the {\em stress response functions} (analogous to the pressure vector for the Oseen problem \cite{Dhont})  are given by 
\begin{align} 
\label{eq:sigmaxf}
\wchi_{xz}(z; \omega) & = \rho_0\nus \partial_z {\widetilde R}_x(z; \omega),\\
\label{eq:sigmazf}
\wchi_{zz}(z; \omega) & = \rho_0\nun \left( 1 + {\icomplex} \frac{c_0^2}{\nun\omega}\right) \partial_z {\widetilde R}_z(z; \omega). 
\end{align}

Equations \eqref{eq:sigmaxf1}-\eqref{eq:sigmazf} can be used to evaluate the desired two-point correlators of the stresses across the fluid film ($0\leq z, z'\leq h$) defined as 
\begin{equation}
\label{eq:csheardef_0}
{\mathcal C}_{jz} (z,z'; t-t') = \langle \sigma_{jz}(z;t)\sigma_{jz}(z';t') \rangle,
\end{equation}
for $ j=x,z$, where the time homogeneity of the correlators in the stationary state is also explicitly indicated. 
The average in Eq. \eqref{eq:csheardef_0} can be evaluated (Section \ref{subsec:EOM}), yielding  the transverse/longitudinal stress correlators in the frequency domain in terms of the stress response functions \eqref{eq:sigmaxf} and \eqref{eq:sigmazf} as
\begin{equation} 
\label{eq:sigmaxcorr}
{\widetilde{\mathcal C}}_{jz} (z,z'; \omega) = 2{\widetilde {\mathcal G}_j}(\omega)\wchi_{jz}(z; \omega)  \wchi_{jz}(z'; -\omega). 
\end{equation}
 Other relevant quantities include the two-point correlators of density and pressure fluctuations  expressed as 
\begin{align}
\label{eq:density_corr_def}
&{\mathcal C}_{\rho\rho} (z,z'; t-t') = \langle \rho(z;t)\rho(z';t') \rangle,  \\
\label{eq:pressure_corr_def}
&{\mathcal C}_{pp} (z,z'; t-t') = \langle p(z;t)p(z';t') \rangle.  
\end{align}
These quantities are related through ${\mathcal C}_{pp} = {\mathcal C}_{\rho\rho}/c_0^4$ (see  Eq. \eqref{eq:prho}). In the present context, the frequency-domain forms (or, the spectral densities) of  
density correlator, ${\widetilde{\mathcal C}}_{\rho\rho} (z,z'; \omega)$, and the corresponding pressure correlator, ${\widetilde{\mathcal C}}_{pp} (z,z'; \omega)$, are found to be related directly to that of the {\em longitudinal} stress correlator as 
\begin{equation}
\label{eq:density_corr}
 {\widetilde{\mathcal C}}_{\rho\rho} (z,z'; \omega) = c_0^{-4}\, {\widetilde{\mathcal C}}_{pp} (z,z'; \omega)  =\frac{{\widetilde{\mathcal C}}_{zz} (z,z'; \omega)}{c_0^4+\nun^2\omega^2}.
\end{equation}
 
The two-point density correlator is of special interest in the context of thermal (near-equilibrium) hydrodynamic fluctuations in bulk fluids, in which case one uses its frequency/wavevector representation to obtain the spectral density of density fluctuations  \cite{Boon-Yip,Berne-Pecora}. 
This latter quantity can be measured through inelastic (polarized) light scattering methods, enabling experimental determination of hydrodynamic-fluctuation spectra as well as various thermodynamic quantities and transport coefficients of bulk fluids. Our formulation directly relates the spectral density of hydrodynamic stress fluctuations on the film boundaries to the spectral density of fluid density fluctuations within the film. As such, it suggests an alternative method to probe the density fluctuations through the measurements of surface forces in confined fluids, where standard light scattering methods may be less suitable.

\subsection{Dimensionless representation}
\label{subsec:dimless}

In our later treatment of shear ($\parallel$) and compressional ($\perp$) modes in dimensionless units, we shall make use of the rescaled variables  
\begin{equation}\label{eq:omegasdef}
\zs = \frac{z}{\nusn/c_0},\quad \taus = \frac{t}{\nusn/c_0^2} \quad \mathrm{and} \quad  \omegas = \frac{\omega}{c_0^2/\nusn}, 
\end{equation}
as well as the rescaled inter-plate separation (rescaled film thickness) and the dimensionless mass per unit area of the mobile plate, respectively, as   
\begin{equation}
\hs=\frac{h}{\nusn/c_0}\qquad\mathrm{and}\qquad\gammas = \frac{m}{\rho_0 \nusn/c_0}.
\end{equation}
Using the definitions in Eqs. \eqref{eq:alphaRI}-\eqref{eq:kappaI}, the parameter $\alpha(\omega)$  can be rescaled as ${\nus}\alpha(\omega)/{c_0}\rightarrow  \xi(\omegas)$, where 
\begin{equation}
\xi(\omegas)  =\pm\sqrt{\frac{|\omegas|}{2}}\left(1+{\icomplex}\sgn(\omegas)\right), 
\label{eq:xidef}
\end{equation}
while  $\kappa(\omega)$ can be rescaled  as $ \nun\kappa(\omega)/c_0\rightarrow \ell(\omegan) = {\ell_\mathrm{R}}(\omegan) + {\icomplex}{\ell_\mathrm{I}}(\omegan)$, with real and imaginary parts 
\begin{align}
{\ell_\mathrm{R}} (\omegan) &
= \pm\frac{|\omegan|}{\sqrt{2}}  \sqrt{\frac{\sqrt{1+\omegan^2}+1}{1+\omegan^2}}, \label{eq:ellRdef}\\
{\ell_\mathrm{I}} (\omegan) &
= \pm\frac{\omegan}{\sqrt{2}} \sqrt{\frac{\sqrt{1+\omegan^2}-1}{1+\omegan^2}}.
\label{eq:ellIdef}
\end{align}
The spectral densities of external surface forcing are rescaled 
as ${\widetilde {\mathcal G}_x}(\omega)/{\mathcal G}_x\rightarrow \wcalGshear(\omegas)$ and ${\widetilde {\mathcal G}_z}(\omega)/{\mathcal G}_z\rightarrow \wcalGnorm(\omegan)$. 

The  transverse and longitudinal stress response functions (\ref{eq:sigmaxf}) and (\ref{eq:sigmazf}) are dimensionless  and can be re-expressed in rescaled coordinates immediately as $\wchi_{xz} (z; \omega)\rightarrow {\wchishear} (\zs; \omegas)$ and $\wchi_{zz} (z; \omega)\rightarrow {\wchinorm} (\zn; \omegan)$ with explicit forms given by
\begin{align}
 \label{eq:chi_s_omega_final}
&{\wchishear} (\zs; \omegas) =  - \frac{\cos[\xi(\omegas)(\hs-\zs)]}{ \cos[\xi(\omegas)\hs] - \gammas \xi(\omegas) \sin[\xi(\omegas)\hs] },  \\
 \label{eq:chi_n_omega_final}
&{\wchinorm} (\zn; \omegan) = - \frac{\cos[\ell(\omegan)(\hn-\zn)]} {\cos[\ell(\omegan)\hn] - \gamman \ell(\omegan) \sin[\ell(\omegan)\hn]}. 
\end{align}

The transverse and longitudinal stress correlators in the time domain, Eq. \eqref{eq:csheardef_0}, are rescaled by the characteristic stress variances  $2{\mathcal G}_x c_0^2 / \nus$ and $2{\mathcal G}_z c_0^2 / \nun$ as
\begin{align}
\label{eq:cshear_def}
\frac{{\mathcal C}_{xz} (z,z'; t-t')}{2{\mathcal G}_x c_0^2 / \nus}&\rightarrow \cshear (\zs,\zs'; \Delta \taus),  \\
\frac{{\mathcal C}_{zz} (z,z'; t-t')}{2{\mathcal G}_z  c_0^2 / \nun}&\rightarrow \cnorm (\zn,\zn'; \Delta \taun), 
\label{eq:cnorm_def}
\end{align}
where we have defined $\Delta \taun = \taun - \taun'$ and $\Delta\taus = \taus-\taus'$. These time-dependent (real-valued) stress correlators are to be calculated from the Fourier-transform relations 
\begin{align}
\label{eq:cshearintegral}
&\cshear(\zs, \zs'; \Delta \taus) = 
\int  \frac{\mathrm{d}\omegas}{2\pi}\, e^{-{\icomplex}\omegas \Delta \taus}\wcshear(\zs, \zs'; \omegas),\\
\label{eq:cnormintegral}
&\cnorm(\zn, \zn'; \Delta \taun) = 
\int \frac{\mathrm{d}\omegan}{2\pi}\, e^{-{\icomplex}\omegan \Delta \taun}\wcnorm(\zn,\zn'; \omegan), 
\end{align}
and also by making use of the frequency-domain expressions for $\wcshear(\zs, \zs'; \omegas)$ and $\wcnorm(\zn,\zn'; \omegan)$, themselves following from the dimensionless form of Eq.  \eqref{eq:sigmaxcorr} as 
\begin{align}
\label{eq:wcshear_nondim}
&\wcshear(\zs, \zs'; \omegas) = \wcalGshear(\omegas){\wchishear}(\zs; \omegas) {\wchishear}(\zs'; -\omegas),\\
\label{eq:wcnorm_nondim}
&\wcnorm(\zn,\zn'; \omegan) = \wcalGnorm(\omegan){\wchinorm}(\zn; \omegan) {\wchinorm}(\zn'; -\omegan).  
\end{align}
Since $\xi(-\omegas) = \xi^\ast(\omegas)$ and $\ell(-\omegan)=\ell^\ast(\omegan)$ (Section \ref{subsec:vel_dens_fields}) and, hence, ${\wchishear}(\zs; -\omegas) = {\wchishear}^\ast(\zs; \omegas)$ and ${\wchinorm}(\zs; -\omegas) = {\wchinorm}^\ast(\zs; \omegas)$, the stress correlators are found to be symmetric w.r.t. interchanging their spatial (first and second)  arguments and concurrently reversing the sign of their frequency/time (third) argument (or, by taking their complex conjugates); i.e., $\wcshear(\zs, \zs'; \omegas) = \wcshear(\zs', \zs; -\omegas) = \wcshear^\ast(\zs', \zs; \omegas)$, and  $\wcnorm(\zn,\zn'; \omegan) = \wcnorm(\zn',\zn; -\omegan) = \wcnorm^\ast(\zn',\zn; \omegan)$. Formally similar symmetry relations hold in the time domain for $\cshear(\zs, \zs'; \Delta \taus)$ and $\cnorm(\zn, \zn'; \Delta \taun)$, and also for the density and pressure correlators as implied by Eq. \eqref{eq:density_corr}. It should also be noted that the {\em same-point} correlators, $\wcshear(\zs, \zs; \omegas)$ and $\wcnorm(\zn,\zn; \omegan)$, 
are nothing but the local {\em variances} of stress fluctuations at a given frequency. They are evidently real-valued and positive and, as such, ensure the stability of our linear-fluctuation analysis. 

For the sake of concreteness in our numerical analysis below, we set the forcing spectral densities equal to one, $\wcalGshear(\omegas)= \wcalGnorm(\omegan)=1$, equivalent to adopting the {\em white-noise Ansatz} for the surface forcing as noted before.

\section{Transverse stress correlators}
\label{sec:shear_corr}

We start our analysis by focusing first on the transverse hydrodynamic stresses produced by the white noise external surface forcing. The two-point correlator in the frequency domain, $\wcshear(\zs, \zs'; \omegas)$, can be shown to be only a function of the redefined dimensionless frequency and coordinate variables $\omegas\hs^2$ and $\zs/\hs$ and the rescaled parameter  $\hs/\gammas$. By setting $\zs, \zs'=\{0, \hs\}$, we obtain the {\em same-plate} and {\em cross-plate} correlators, which we shall refer to as {\em self-} and {\em cross-correlators}, respectively, as 
 \begin{align}
 &\wcshear(0, 0; \omegas)=\vert{\wchishear}(0; \omegas)\vert^2,\,\,\,
 \wcshear(\hs, \hs; \omegas)= \vert{\wchishear}(\hs; \omegas)\vert^2, 
 \nonumber \\
  \label{eq:sameplate_shear_w_cross}
 &\wcshear (0,\hs; \omegas) = \wcshear^\ast (\hs, 0; \omegas) ={\wchishear}(0; \omegas){\wchishear}^\ast(\hs; \omegas). 
\end{align}
In Fig. \ref{fig:spectra_shear}, we show the self-correlators, $\wcshear(0, 0; \omegas)$  and $\wcshear(\hs, \hs; \omegas)$, and the real part of the cross-correlator, $\real \, \wcshear (0,\hs; \omegas)$ (being the only part that contributes to the cross-correlator in the time domain) as functions of  $\omegas\hs^2$ 
for $\hs/\gammas=5$. 
As seen, the three quantities exhibit rapid monotonic decrease with the frequency from their maximum value of one at zero frequency fulfilling the inequalities $\real \, \wcshear (0,\hs; \omegas)\leq\wcshear(\hs, \hs; \omegas)\leq \wcshear(0, 0; \omegas)$ across the frequency domain.

\begin{figure}[t!]
\begin{center}  
\includegraphics[width=5.65cm]{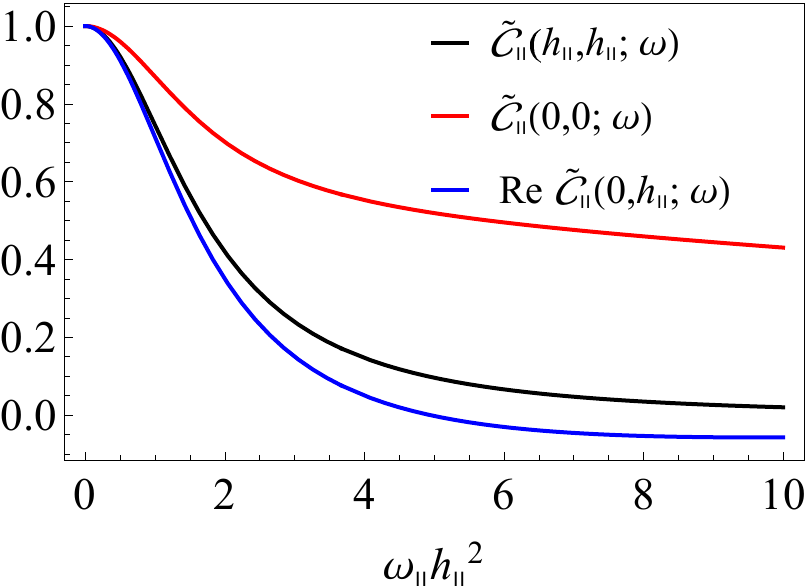} 
\caption{The self- and cross-correlators of the transverse stresses acting on the plates as functions of the redefined dimensionless frequency, $\omegas\hs^2$, for fixed $\hs/\gammas=5$. 
} 
\label{fig:spectra_shear}
\end{center}
\end{figure}

\begin{figure}[t!]
\begin{center}  
\includegraphics[width=6.cm]{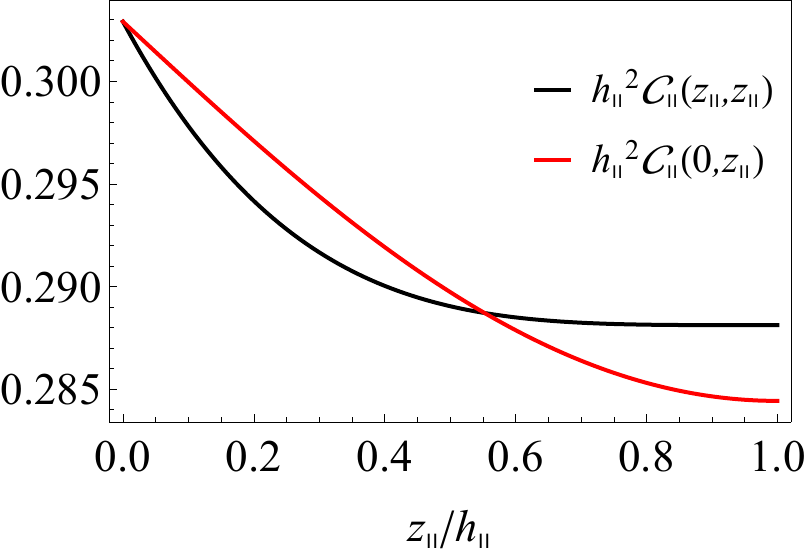} 
\caption{Profiles of the rescaled, equal-time, transverse stress correlators $\hs^2\cshear (\zs,\zs)$ and $\hs^2\cshear (0,\zs)$ across the fluid film ($0\leq \zs\leq \hs$) for fixed $\hs/\gammas=5$. 
} 
\label{fig:stress_profiles_shear}
\end{center}
\end{figure}

\begin{figure*}[t!]
\centering
\includegraphics[width=5.65cm]{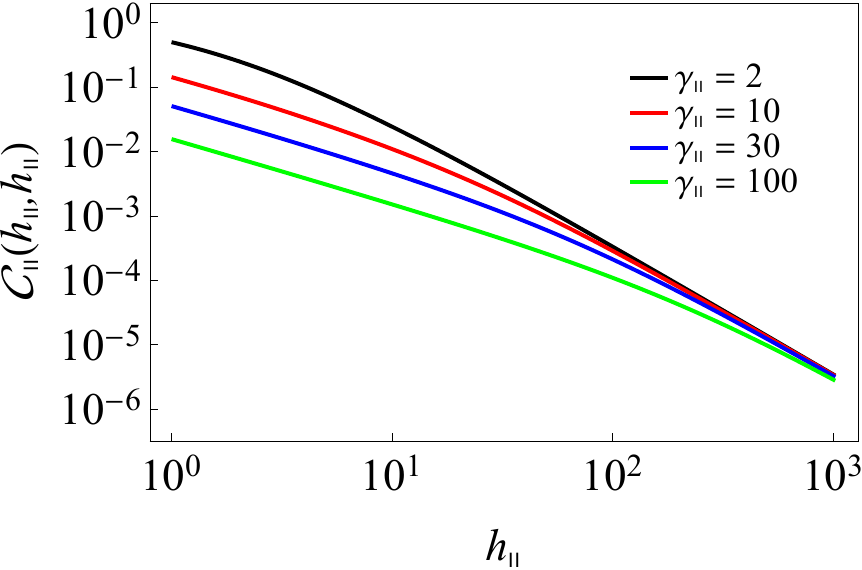} \hspace{-5mm} (a)
\includegraphics[width=5.75cm]{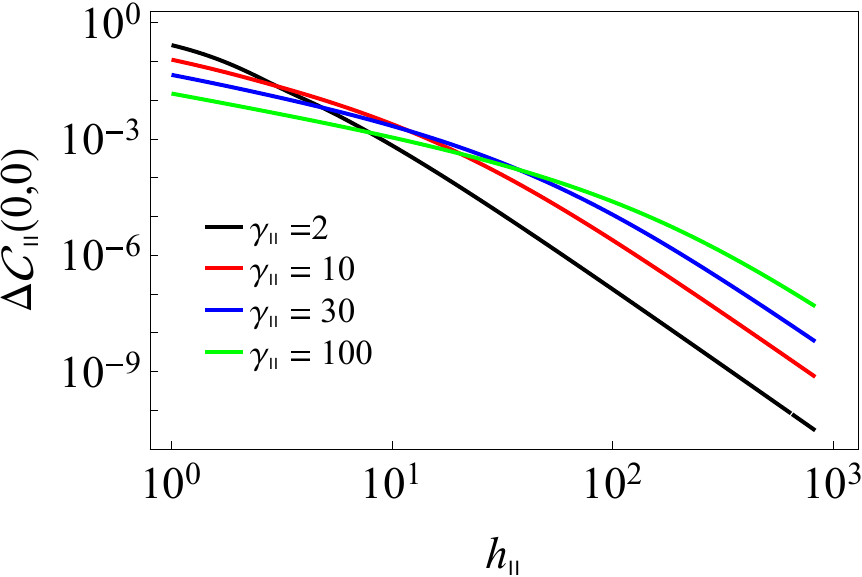} \hspace{-5mm} (b)
\includegraphics[width=5.65cm]{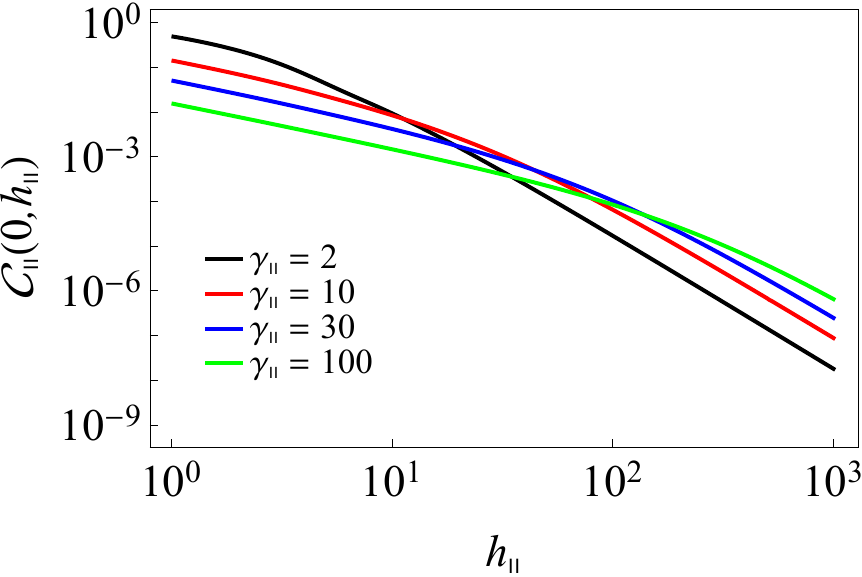} \hspace{-5mm} (c)
\caption{Log-log plots of equal-time  transverse stress (self- and cross-) correlators $\cshear(\hs, \hs)$, $\Delta\cshear(0, 0)$ and $\cshear(0, \hs)$, in panels (a), (b) and (c), respectively, as functions of the dimensionless inter-plate separation, $\hs$, for $\gammas=2, 10, 30$ and $100$.
\label{fig:equaltimeCs}
}
\end{figure*} 

These behaviors can be understood by noting that the characteristic frequencies of the transverse modes occur at the poles of the transverse response function ${\wchishear}$  
over the complex frequency ($\varsigma$) plane. These poles are found to fall onto the lower-half imaginary axis, reflecting the diffusive nature of the shear  modes in the hydrodynamic regime   (Appendix \ref{app:poles}). The singular part of the transverse response functions in the proximity of a given pole $\varsigma_{\!_\parallel}^{(n)}$ (see Eq. \eqref{eq:sol_roots_shear_app2})  behaves as  ${\wchishear}\sim 1/\big(\omegas^2+[\imaginary\,\varsigma-\imaginary\,\varsigma_{\!_\parallel}^{(n)}]^2\big)$. Thus, along the real-valued frequency axis $\omegas = \real\,{\varsigma}$, ${\wchishear}$ takes a Lorentzian form peaked around $\omegas=0$ as all of the poles are purely imaginary, explaining the monotonically decreasing behaviors  in Fig. \ref{fig:spectra_shear}.  

\subsection{Equal-time transverse stress correlators}
\label{subsec:shear_corr}
 
The equal-time transverse correlators are obtained by setting $\Delta \taus=0$ in Eq. \eqref{eq:cshearintegral}, in which case we  drop the time (third) argument by redefining our notation as $\cshear (\zs,\zs') \equiv\cshear (\zs,\zs'; \Delta \taus=0)$. Hence, 
\begin{align}
\label{eq:shear_equal_time_redef}
\cshear (\zs,\zs') =  \int  \frac{\mathrm{d}\omegas}{2\pi}\,{\wchishear}(\zs; \omegas){\wchishear}^\ast(\zs'; \omegas). 
\end{align}
When rescaled with the inter-plate separation as $\hs^2 \cshear (\zs,\zs)$ and $\hs^2 \cshear (0,\zs)$, these correlators will be functions of $\zs/\hs$ and $\hs/\gammas$ only. As seen in Fig. \ref{fig:stress_profiles_shear}, both of these correlators vary within a relatively small range of values across the film. Although this may naively appear as indicating an approximate power-law behavior of $\cshear \sim \hs^{-2}$ at any point across the film, we find a more diverse set of power-law behaviors for the self- and cross-correlators of the stresses acting on the plates.

\begin{figure}[t!]
\begin{center}  
\includegraphics[width=5.75cm]{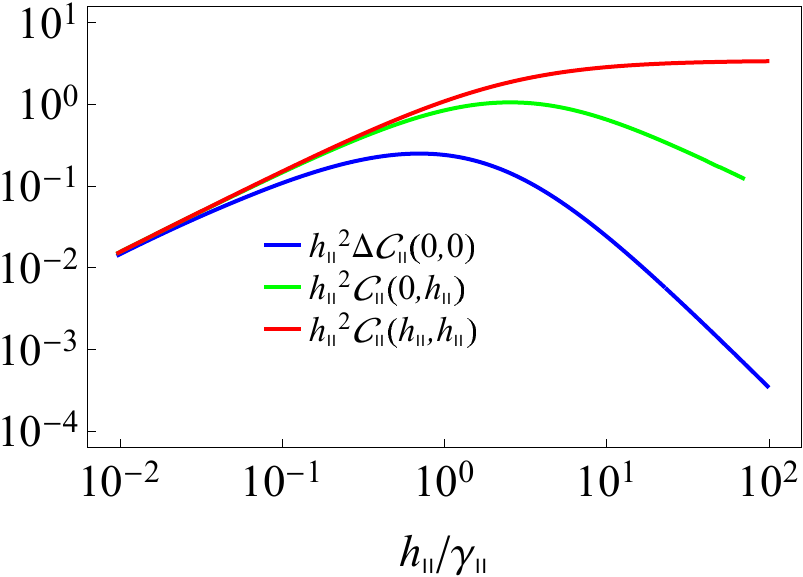} 
\caption{The rescaled universal forms of the three equal-time transverse stress correlators, as indicated on the graph (see also Figs. \ref{fig:equaltimeCs}a-c), as functions of $\hs/\gammas$,  demonstrating the crossover between the two power-law regimes found at small and large values of $\hs/\gammas$ in each case. 
} 
\label{fig:stress_profiles_univ}
\end{center}
\end{figure}

It should be noted that the frequency integrals discussed above converge sufficiently rapidly and, therefore, remain  finite. For numerical expediency, and as a check on the self-consistency of the continuum approach used here (see Appendix \ref{app:cutoff} for details), it is convenient to introduce a high frequency cutoff $\omegas^\infty$, which (as discussed in the appendix) can conveniently be taken as $\omegas^\infty=1$. In general, the  numerical outcomes for $\cshear(\hs, \hs)$ and $\cshear(0, \hs)$ remain unaffected by the precise choice of the frequency cutoff, when the latter is sufficiently large.  
The self-correlator on the {\em mobile} (lower) plate, $\cshear(0, 0)$, however tends to a  constant predominantly determined by the forcing spectral density as the inter-plate separation is increased.
To eliminate such spurious effects, we subtract this limiting value that  represents the self-interaction of the mobile plate in the bulk by defining the {\em excess} correlator as 
$\Delta\cshear(0, 0) = \cshear(0, 0) - \lim_{\hs\rightarrow\infty} \cshear(0, 0)$.  
In all cases, our numerical results remain independent of the choice of the frequency cutoff, when the latter is large enough, but it is still kept within the regime consistent with the continuum hydrodynamic description (Appendix \ref{app:cutoff}).

\begin{figure*}[t!]
\centering
\includegraphics[width=5.8cm]{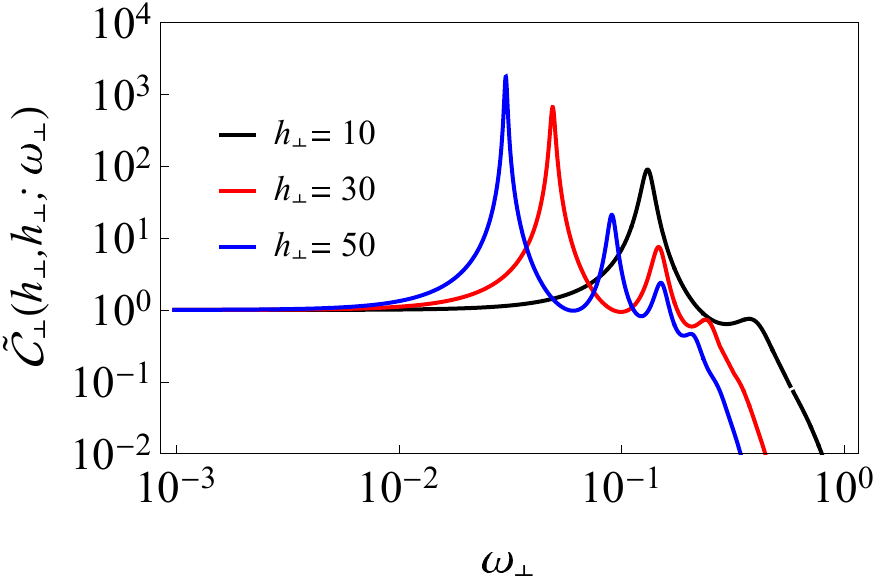} \hspace{-5mm} (a)
\includegraphics[width=5.55cm]{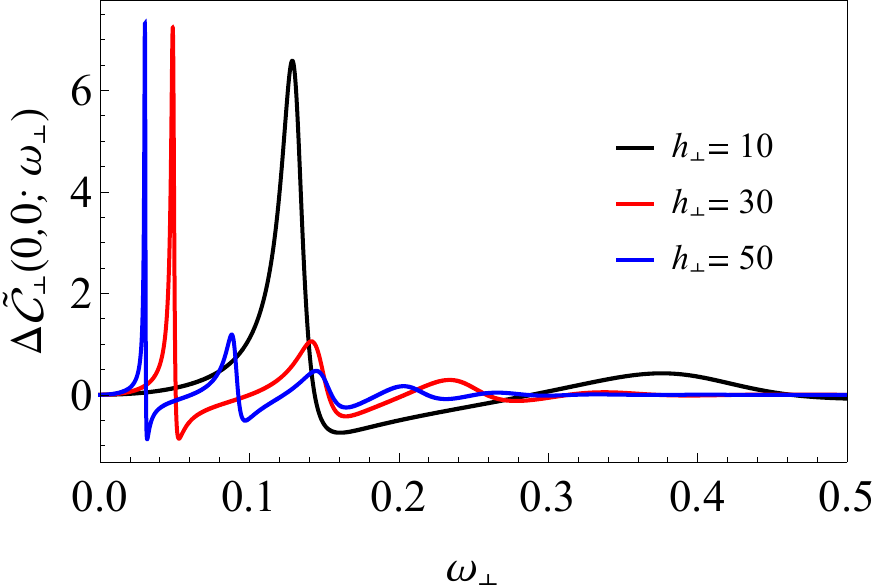} \hspace{-5mm} (b)
\includegraphics[width=5.85cm]{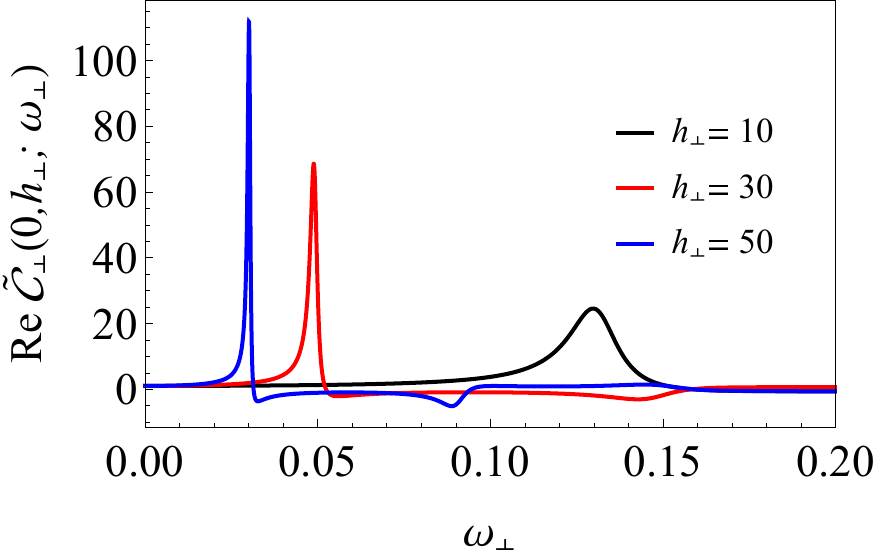} \hspace{-5mm} (c)
\caption{Log-log plot of the longitudinal stress self-correlator $\wcnorm(\hn,\hn; \omegan)$ in panel (a), and linear plots of the excess longitudinal stress (self- and cross-) correlators $\Delta \wcnorm(0,0; \omegan)$ and $\real\,\wcnorm(0,\hn; \omegan)$ in panels (b) and (c), respectively, as functions of the dimensionless frequency, $\omegan$, for fixed $\gamman=2$ and $\hn=10, 30$ and $50$ as indicated on the graphs. 
\label{fig:Cn_freq}
}
\end{figure*} 

As the log-log plots in Fig. \ref{fig:equaltimeCs} show, the stress correlators fall off rapidly as power-laws of the rescaled inter-plate separation, $\hs$, in all cases (panels a to c), when $\hs$ is sufficiently large. In fact, one can discern distinct power-law regimes for both small and large values of $\hs$ as $\gammas$ is varied. As noted before, the key parameter here is the ratio $\hs/\gammas$. We find the power-law behaviors as  
\begin{align}
\label{eq:powerlaw_s_hh}
\cshear(\hs, \hs)\sim\left\{\begin{array}{l l}
  \hs^{-2} & \quad \hs/\gammas\gg 1, \\   
    \hs^{-1} & \quad \hs/\gammas\ll 1,     
\end{array}\right.
\end{align}
\begin{align}
\label{eq:powerlaw_s_00}
\Delta\cshear(0, 0)\sim\left\{\begin{array}{l l}
  \hs^{-4} & \quad \hs/\gammas\gg 1, \\   
    \hs^{-1} & \quad \hs/\gammas\ll 1,    
\end{array}\right.
\end{align}
\begin{align}
\label{eq:powerlaw_s_0h}
\cshear(0, \hs)\sim\left\{\begin{array}{l l}
  \hs^{-3} & \quad \hs/\gammas\gg 1, \\   
    \hs^{-1} & \quad \hs/\gammas\ll 1. 
\end{array}\right.
\end{align}
Because of the cumbersome form of the integrand in Eq. \eqref{eq:cshearintegral} (involving factors with implicit and complex-valued dependence on $\omegas$), we were not able to derive the scaling forms analytically.  They have rather been confirmed numerically, with the reported {\em universal} exponents being accurate within a margin of error $<5\%$. These power-laws reflect the more fundamental scale-invariant forms admitted in general by each of the above transverse stress correlators in terms of their two main parameters as
${\mathcal C}_{\!_\parallel} = \hs^{-2}{\mathcal F}\left({\gammas}/{\hs}\right)$, 
where ${\mathcal C}_{\!_\parallel}$ stands for either of the quantities $\cshear(\hs, \hs)$, $\Delta\cshear(0, 0)$ or $\cshear(0, \hs)$, and ${\mathcal F}(\cdot)$ is the corresponding {\em universal} function.  For these three cases, the universal functions can be calculated numerically for a wide range (several decades) of their arguments with the results, shown in the log-log plot of Fig. \ref{fig:stress_profiles_univ}, clearly demonstrating the crossover between two distinct power-law regimes: These regimes appear  as straight lines for small and large $\hs/\gammas$ for each one of the three plotted curves.  The numerical values of the slopes match the corresponding exponents given in Eqs. \eqref{eq:powerlaw_s_hh}-\eqref{eq:powerlaw_s_0h}  plus two. 

\section{Longitudinal stress correlators}
\label{sec:norm_corr}

The longitudinal stress correlators show distinct features as compared with their transverse counterparts. As seen from the frequency-domain plots in panels a to c in Fig. \ref{fig:Cn_freq}, the longitudinal self- and cross-correlators exhibit well-developed peaks, representing acoustic resonances due to the compressional modes excited by the external surface forcing in the fluid film. These modes are associated with the poles of the corresponding response functions in the complex frequency plane (see Appendix \ref{app:poles}).  The peaks observed along the real-frequency axis, $\omegan$, in Fig. \ref{fig:Cn_freq} are indeed produced by the first few compression poles (for instance, the four peaks seen for $\hn=50$, blue curve, in $\wcnorm(\hn,\hn; \omegan)$, panel a, coincide with the loci of the first four poles with $0\leq n \leq 3$). 
The number of peaks and their heights grow and their loci shift to smaller frequencies as the inter-plate separation, $\hn$, is increased (and/or as $\gamman$ is increased). At a given inter-plate separation, the longitudinal stress correlators take their largest values at the first peak, which gives the dominant contribution to the frequency integrals. This signifies the prevalence of low-frequency acoustic resonances at larger inter-plate separations and the major role of the corresponding acoustic modes propagating across the film in intensifying fluctuations and correlations of the longitudinal stresses exerted on the confining plates (the higher-order acoustic modes are more strongly attenuated as higher-order peak heights are suppressed by one or even few orders of magnitude as seen in panel a).

\begin{figure}[t!]
\begin{center}  
\includegraphics[width=5.5cm]{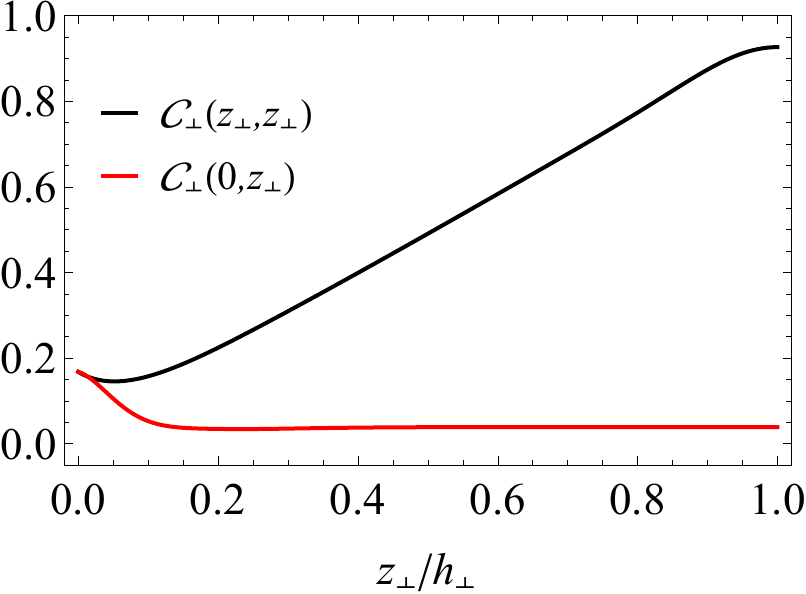} 
\caption{Profiles of the equal-time longitudinal stress correlators $\cnorm (\zn,\zn)$ and $\cnorm (0,\zn)$ across the fluid film ($0\leq \zn\leq \hn$) for fixed $\hn=50$ and $\gamman=2$. 
} 
\label{fig:stress_profiles_norm}
\end{center}
\end{figure}

\begin{figure*}[t!]
\centering
\includegraphics[width=5.65cm]{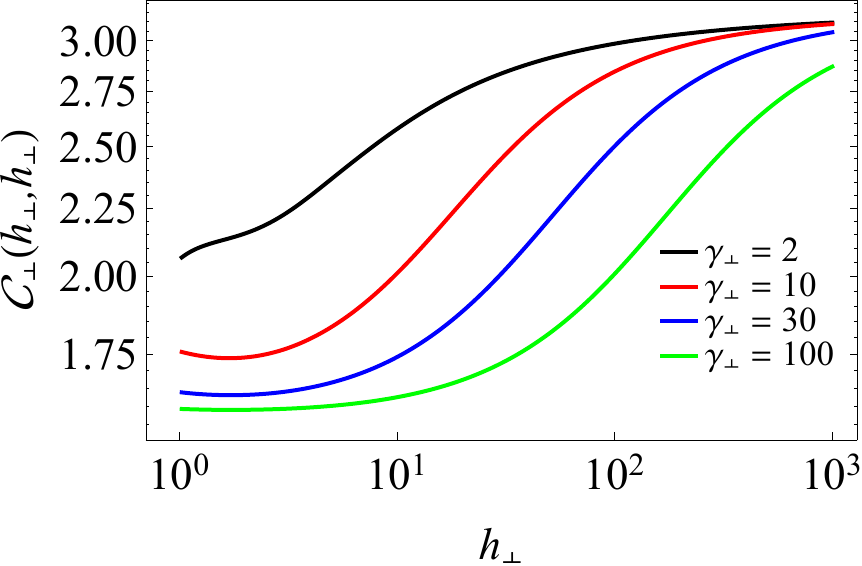} \hspace{-5mm} (a)
\includegraphics[width=5.65cm]{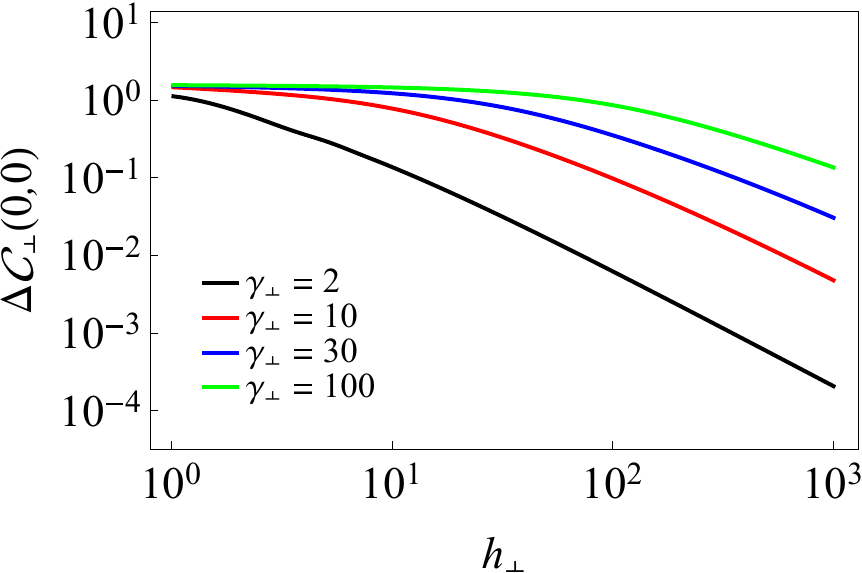} \hspace{-5mm} (b)
\includegraphics[width=5.65cm]{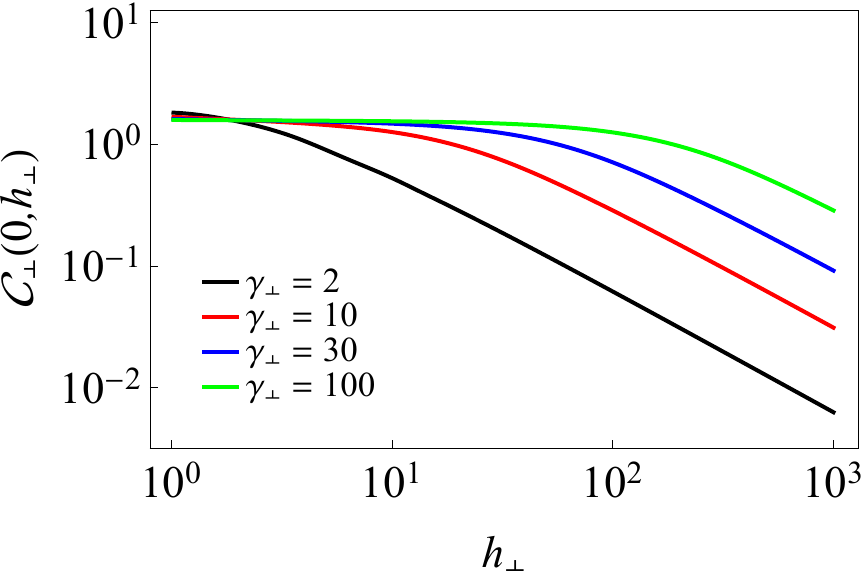} \hspace{-5mm} (c)
\caption{Log-log plots of equal-time  longitudinal stress (self- and cross-) correlators $\cnorm(\hn, \hn)$, $\Delta\cnorm(0, 0)$ and $\cnorm(0, \hn)$, in panels (a), (b) and (c), respectively, as functions of the dimensionless inter-plate separation, $\hn$, for $\gamman=2, 10, 30$ and $100$.
\label{fig:equaltimeCn}
}
\end{figure*} 

Although, such acoustic resonances strengthen the longitudinal stress correlator on the upper (fixed) plate, $\wcnorm(\hn,\hn; \omegan)$, which thus remains consistently above its zero-frequency value of one for a wide range of frequencies, sound absorption becomes gradually dominant (as the imaginary parts of the compression poles become large; Appendix \ref{app:poles}), causing the stress correlator to fall off to zero at sufficiently high frequencies (for $\hn=50$ in panel a, this occurs on approach, but well before, the chosen upper frequency cutoff of  $\omegan^\infty=1$; see Appendix \ref{app:cutoff}). In the case of $\Delta\wcnorm(0,0; \omegan)$ and $\wcnorm(0,\hn; \omegan)$ in panels b and c in Fig. \ref{fig:Cn_freq}, the correlators can take both positive and negative values. The negative values of these quantities represent out-of-phase (or anti-) correlations occurring in certain intervals along the real-frequency axis.

\subsection{Equal-time longitudinal stress correlators}
\label{subsec:normal_corr}

The equal-time longitudinal stress correlators defined through Eq. \eqref{eq:cnormintegral}, and denoted more compactly as $\cshear (\zn,\zn') \equiv\cshear (\zn,\zn'; \Delta \taun=0)$, can be evaluated from 
\begin{align}
\label{eq:norm_equal_time_redef}
\cnorm (\zn,\zn') =  \int  \frac{\mathrm{d}\omegan}{2\pi}\,{\wchinorm}(\zn; \omegan){\wchinorm}^\ast(\zn'; \omegan). 
\end{align}
Unlike their transverse counterparts, these correlators do not in general admit scale-invariant forms. In Fig. \ref{fig:stress_profiles_norm}, we show $\cnorm (\zn,\zn)$ and $\cnorm (0,\zn)$ as functions of $\zn/\hn$ for fixed $\hn=50$ and $\gamman=2$. It is interesting to note that while the different-point ($\zn\neq \zn'$) correlator levels off rapidly to a limiting value smaller than the reference value of $\cnorm (0,0)$, the same-point correlator of longitudinal stresses increases almost linearly as one moves away from the lower (mobile) plate toward the upper (fixed) plate, where it takes its largest value. This indicates stronger hydrodynamic stress fluctuations closer to the fixed plate and pronounced, {\em non-decaying}, r.m.s. values for the longitudinal   stresses acting on it even at relatively large separations. 

This non-decaying behavior is more clearly seen from the plot in panel a in Fig. \ref{fig:equaltimeCn}. Here, the equal-time longitudinal correlator, $\cnorm(\hn,\hn)$, is shown to increase to a saturated maximum level, depending on the dimensionless mass parameter $\gamman$. The reason for this behavior is that as $\hn$ is gradually increased (starting from its minimum base value of $\hn=1$), the characteristic frequency corresponding to the first compressional mode decreases and falls below the cutoff frequency of $\omegan^\infty=1$, making it realizable and relevant in the hydrodynamic domain, manifesting itself also as a gradual increase in $\cnorm(\hn,\hn)$. As $\hn$ is further increased, the first characteristic frequency (whose locus over the real-frequency axis scales as $\hn^{-1}$, as is characteristic to acoustic modes) decreases further toward the low-frequency regions, where sound attenuation is subdominant. As such, the corresponding acoustic resonance leads to a strongly propagating acoustic mode across the fluid film, creating a pronounced first peak in the stress-correlator plots in the frequency domain (panel a in Fig. \ref{fig:Cn_freq}), bringing $\cnorm (\hn,\hn)$ up to a saturation level (panel a in Fig. \ref{fig:equaltimeCn}; see also panel b in Fig. \ref{fig:cutoff_n} in Appendix \ref{app:cutoff}). The crossover to the saturation regime in $\cnorm (\hn,\hn)$  occurs roughly at $\hn\sim \gamman$. 

In the case of the longitudinal stress correlators $\Delta\cnorm(0, 0)$ and $\cnorm(0, \hn)$,  shown in panels b and c, and in analogy with our discussion of the power-law behaviors in the case of transverse correlators, we find power-law behaviors in the regime of large inter-plate separations, $\hn/\gamman\gg 1$, with universal exponents as 
\begin{align}
&\Delta\cnorm(0, 0)\sim \hn^{-3/2}, \\   
&\cnorm(0, \hn)\sim\hn^{-1}.
\end{align}
The excess stress correlator on the lower (mobile) plate is defined here as $\Delta\cnorm(0, 0) = \cnorm(0, 0) - \lim_{\hn\rightarrow\infty} \cnorm(0, 0)$. 
 
In all cases discussed above, the main contribution to the longitudinal correlators comes from the fluctuations and correlation produced in the film through the pressure (second) term in Eq. \eqref{eq:sigmaz_0} rather than the viscous stress (first) terms. 

\section{Discussion and conclusion}
\label{sec:conclusion}

We have studied hydrodynamic correlations and fluctuation-induced interactions mediated between the no-slip bounding surfaces of a planar film of a  compressible and viscous fluid, driven externally at one of its boundaries (lower plate) by a stochastic surface forcing of arbitrary spectral density, while the other (upper) plate is kept fixed. We develop general analytical results within the linear hydrodynamic scheme and numerically analyze the outcomes for the special case of a Gaussian white-noise forcing. The stochastic surface forcing  leads to fluctuating transverse (shear) and longitudinal (compressional) hydrodynamic stresses within the film and on the bounding surfaces. To bring out the hydrodynamic fluctuation-induced effects more clearly, we conveniently assume that the external forcing has a zero mean; hence, the resulting hydrodynamic stresses also vanish on average, and their two-point correlators embody the hydrodynamic correlation effects.

We show that the same-plate (self-) and cross-plate (cross-) correlators of the transverse stress exhibit two distinct regimes of power-law behaviors at small and  large inter-plate separations $h$, with different, yet universal, scaling exponents. In the case of longitudinal stress correlators, the power-law dependencies are obtained only for the large-separation behavior of the (excess) self-correlator on the mobile plate, $\Delta\cnorm (0,0)$,  and the cross-correlator, $\cnorm (0,h)$, with power-law decays being expressively weaker than those of the transverse correlators. The spectral analysis of the longitudinal stress correlators reveals distinct underlying differences with the transverse ones due to the prevalence of propagating, underdamped, acoustic modes in the confined geometry \cite{Diamant}. 
The longitudinal stress self-correlator at the {\em fixed} plate, $\cnorm (h,h)$,  thus displays a thoroughly different  behavior: It increases with  $h$ and saturates at a finite value, representing a constant, longitudinal,   r.m.s.  stress  $\sigma^{\mathrm{rms}}_{\!_{\perp}}= \sqrt{\cnorm (h,h)}$ at sufficiently large $h$. This feature indicates the existence of {\em non-decaying}, longitudinal, hydrodynamic fluctuation forces, acting on the fixed plate. This can be contrasted with, e.g., the transverse stresses on the fixed plate, whose r.m.s. decays with the inter-plate separation as  $\sigma^{\mathrm{rms}}_{\!_{\parallel}}\sim h^{-1}$ for large $h$. 

The non-decaying behavior of $\cnorm (h,h)$  emanates directly from the excitation of the acoustic modes (acoustic resonances) and sets in at the appearance of the first peak in the corresponding spectral representation  (Section \ref{sec:norm_corr}). 
Indeed, one can directly verify that the dominant contribution to $\cnorm (h,h)$ comes from the compressional term in the longitudinal stress (second term in Eq. \eqref{eq:sigmaz_0}). 
Such long-ranged, sound-mediated, hydrodynamic correlations have also been found in the context of the correlations between Brownian particles (colloids) in strongly confined quasi-one/two-dimensional geometries (see Ref. \cite{Diamant}  and references therein; see also Ref. \cite{Keyser} for recent experiments on non-decaying colloid-colloid  correlations based on the displacement of the intervening fluid column between the colloids in narrow channels). It should be further noted that our analysis is focused on the stationary-state behavior of the system, implying that the film thickness is traversed by recurring propagations of underdamped sound waves of varying (random) amplitude, continually excited by the longitudinal component of the external forcing applied to the mobile plate. This process generates and maintains a finite and stationary  (non-decaying) compressional r.m.s. stress on the fixed plate.  

The power-law relations obtained here for the (generally non-thermal) surface-driven stress correlators assimilate to the ``secondary" hydrodynamic Casimir-like forces, analyzed previously in the context of thermal fluctuations in a fluid film with fixed boundaries \cite{Monahan}. Such non-equilibrium  fluctuation-induced, or Casimir-like effects have been of considerable interest in other soft-matter contexts in recent years \cite{Jones,Chan,Monahan,Bartolo,Dean,Ajay1,Antezza,Kruger,Kirkpatrick13,Kirkpatrick14,Kafri-Kardar}. It is also important to note that the r.m.s. of the fluctuation-induced forces predicted here exhibits a stronger long-ranged character than the standard electromagnetic vdW-Casimir forces \cite{Kardar,Woods}. In the case of the longitudinal  stresses on the fixed plate, our results (panel a of Fig. \ref{fig:equaltimeCn}) give a rescaled  r.m.s. stress  of the order one over the range of rescaled separations $\hn\sim 10-10^2$ for a wide range of values for $\gamman$. Thus, for the parameter values relevant to water (see Appendix \ref{app:cutoff}), we predict a longitudinal r.m.s. stress (pressure) of the order of the r.m.s. surface forcing, $\sigma^{\mathrm{rms}}_{\!_{\perp}}/\sqrt{\sigma_{f}}\sim 1$, over the range of inter-plate separations, or film thicknesses, $h\sim 2.6\times(10-10^2)$~nm. Here, $\sigma_{f}$ stands for the surface forcing variance in the time domain (related to the forcing spectral density as $\sigma_{f}=2{\mathcal G}_z  \omega^\infty$; see Eq.  \eqref{eq:cnorm_def}, Section \ref{subsec:EOM} and Appendix \ref{app:cutoff}). Being an externally controlled quantity,  $\sigma_{f}$  can be  adjusted  arbitrarily to obtain the experimental resolution required for the verification of our predictions. The force-measurement precision  within the dynamic SFA and AFM techniques can be better than $10$~nN and $10$~pN, respectively \cite{Israela,Butt,Korea,Butt-rev}; hence, in the later case (of thermal-noise AFM), even the effects due to ambient thermal noise have been detectable  \cite{Maali2,Butt-rev,Butt,Haviland,Benmouna,Siria2009,Clarke,Sader3}. 

Despite its geometric simplicity, our model captures the essential features of typical  dynamic SFA/AFM setups \cite{Israelachvili_ROPP,Korea,Butt-rev,Haviland,CottinBizonne2008,Leroy2012,Steinberger2008,Wang2017,Maali2,Maali1,Butt-rev,Alcaraz,Benmouna,Clarke,Sader3,Siria2009,Ellis3,Neto,Bocquet2010,Granick1991b,Klein1998,Klein2007,Bureau2010,Berg2003,Grier}  (see Section \ref{sec:model} for further details). While these techniques have widely been used  in the study of the rheological properties of fluid films, substantial focus has been on their utility in high-precision determination of shear/compressional forces produced on the bounding surfaces confining a fluid film, where one of the surfaces is forced in linear or oscillatory motion. In dynamic SFA, the bounding surfaces are usually taken as two apposed, weakly curved, crossed cylinders (sometimes with large flattened contact areas exposed to the intervening fluid \cite{Butt,Israela,Granick1991b,Klein1998,Berg2003,Israelachvili_ROPP,Bureau2010}), or a sphere and a plane, with the radii of curvature  in either case  (of the order of a few mm or cm) being much larger than the film thickness  (in the range of sub-nm to $\mu$m, or larger) \cite{Israelachvili_ROPP,Granick1991b,Klein1998,Klein2007,CottinBizonne2008,Steinberger2008,Bureau2010,Korea,Leroy2012,Wang2017}. In dynamic AFM, wide flat microlevers \cite{Siria2009}, or relatively large, cantilever-mounted spheres (of radius up to tens of $\mu$m) \cite{Butt,Korea,Butt-rev,Haviland,Wang2017,Alcaraz,Benmouna,Maali1,Maali2} are used in forced oscillations or in spontaneous (thermal) stochastic motions next to a planar substrate, probing the local surface properties, the hydrodynamic/viscoelastic properties of the surrounding fluid, and also the hydrodynamic interactions mediated between the probe and the substrate. The typical film geometries and modes of surface motions employed in the aforementioned setups therefore appear well-suited for testing our theoretical predictions. The main assumptions made within our analytical approach (e.g., using small-amplitude oscillations and linearized hydrodynamic treatment) are also directly relevant to the mentioned experimental setups and also agree with previous modeling approaches  \cite{Israelachvili_ROPP,Korea,Butt-rev,Haviland,CottinBizonne2008,Leroy2012,Steinberger2008,Wang2017,Maali2,Maali1,Butt-rev,Alcaraz,Benmouna,Clarke,Sader3,Siria2009}. To the best our knowledge, however,  the surface-force  experiments have so far been focused merely on the net force mediated between the fluid film boundaries as opposed to force fluctuations and correlations. The predicted behaviors for these latter quantities can be examined by scrutinizing the readily available r.m.s. of the experimentally detected forces as a function of the film thickness. 

Our work thus lays out a self-consistent and systematic hydrodynamic-fluctuations approach, incorporating the often-ignored finite compressibility of the fluid film, which is important in the understanding of the sound-mediated effects. It also places the study of hydrodynamic surface forces induced by externally driven thin films within the newly emerging context of non-equilibrium Casimir-type phenomena. 

By connecting the spectral density of hydrodynamic stresses acting on the surface boundaries and the spectral density of fluid density fluctuations in the film (see Eq. \eqref{eq:density_corr}), our analysis also suggests an alternative method to probe the density fluctuations through the measurements of surface forces in confined fluid films, where standard light scattering methods may be less suitable.

Finally, we note that our numerical results are given here only for the case of white-noise forcing. Other examples of external forcing (such as colored or $1/f$ noises) can  straightforwardly be analyzed using the more general analytical formulas presented  (Section \ref{sec:response}). Other possible avenues that can be explored within the present context  include the  surface curvature effects, the fluid thermal conductivity \cite{Chris2017} and possibly also the role of nonlinear effects such as viscous dissipation \cite{Klein2007}.

\section{Acknowledgements}
 
M.M.-A. and S.M. acknowledge funding from the School of Physics, Institute for Research in Fundamental Sciences (IPM), where the research leading to this publication was performed. A.N. acknowledges partial support from Iran Science Elites Federation (ISEF) and the Associateship Scheme of The Abdus Salam International Centre for Theoretical Physics (Trieste, Italy). R.P. gratefully acknowledges support from the ``1000 Talents Program" of China and from the University of Chinese Academy of Sciences (UCAS).

\appendix


\section{Hydrodynamic modes}
\label{app:poles}

The transverse (longitudinal) hydrodynamic modes are given by the poles of the transverse (longitudinal) response functions, occurring only in the lower half of the complex frequency ($\varsigma$) plane, at the roots of the equation, 
\begin{equation}
\cos[\Upsilon(\varsigma)] - \Theta  \Upsilon(\varsigma) \sin[\Upsilon(\varsigma)] = 0,  
\label{roots_eq_app}
\end{equation}
where we have used Eqs. \eqref{eq:chi_s_omega_final} and \eqref{eq:chi_n_omega_final} and the definitions 
\begin{align}
\label{eq:defs_for_roots_app}
\left\{\begin{array}{l l}
\Theta = \gammas/\hs,\,\, \Upsilon(\varsigma) = \xi(\varsigma)\hs  &: \,\,\, {\textrm{(shear),}} \\ 
\Theta = \gamman/\hn,\,\, \Upsilon(\varsigma) = \ell(\varsigma)\hn  &: \,\,\, {\textrm{(compression).}}
\end{array}\right.
\end{align}
The parameter $\Upsilon(\varsigma)$ can be expressed in terms of its real and imaginary parts as $\Upsilon(\varsigma)=\Upsilon_{\mathrm{R}}(\varsigma)+\icomplex\Upsilon_{\mathrm{I}}(\varsigma)$. 

For shear modes, Eq. \eqref{roots_eq_app} admits a series of roots only on the imaginary axis,
whose loci can be estimated analytically by looking at two limiting cases. For $\Theta\gg 1$ (large dimensionless mass parameter for the mobile plate, $\gammas$, or small dimensionless inter-plate separation, $\hs$) and, on the leading order, Eq. \eqref{roots_eq_app} takes the approximate form  $\exp[2\icomplex\Upsilon(\varsigma)] \simeq 1$ or, equivalently, $\Upsilon_{\mathrm{R}}(\varsigma)\simeq n\pi$ and $\Upsilon_{\mathrm{I}}(\varsigma)\simeq 0$, where $n$ is an integer. For $\Theta\ll 1$, we find the approximate leading-order equations $\Upsilon_{\mathrm{R}}(\varsigma)\simeq (n+1/2)\pi$ and $\Upsilon_{\mathrm{I}}(\varsigma)\simeq 0$. Using the definitions in Eqs. \eqref{eq:xidef} and \eqref{eq:defs_for_roots_app}, the shear modes are located as 
\begin{equation}
\label{eq:sol_roots_shear_app2}
 \varsigma_{\!_\parallel}^{(n)} \simeq -\icomplex \frac{\pi^2}{\hs^2}\times 
\left\{\begin{array}{l l}
(n+1/2)^2&\,\,\, \hs/\gammas\gg 1, \\ 
n^2\,\, (n\neq 0)\,\,&\,\,\, \hs/\gammas\ll 1.
\end{array}\right.
\end{equation}

\begin{figure}[t!]
\begin{center}  		
\includegraphics[width=5.65cm]{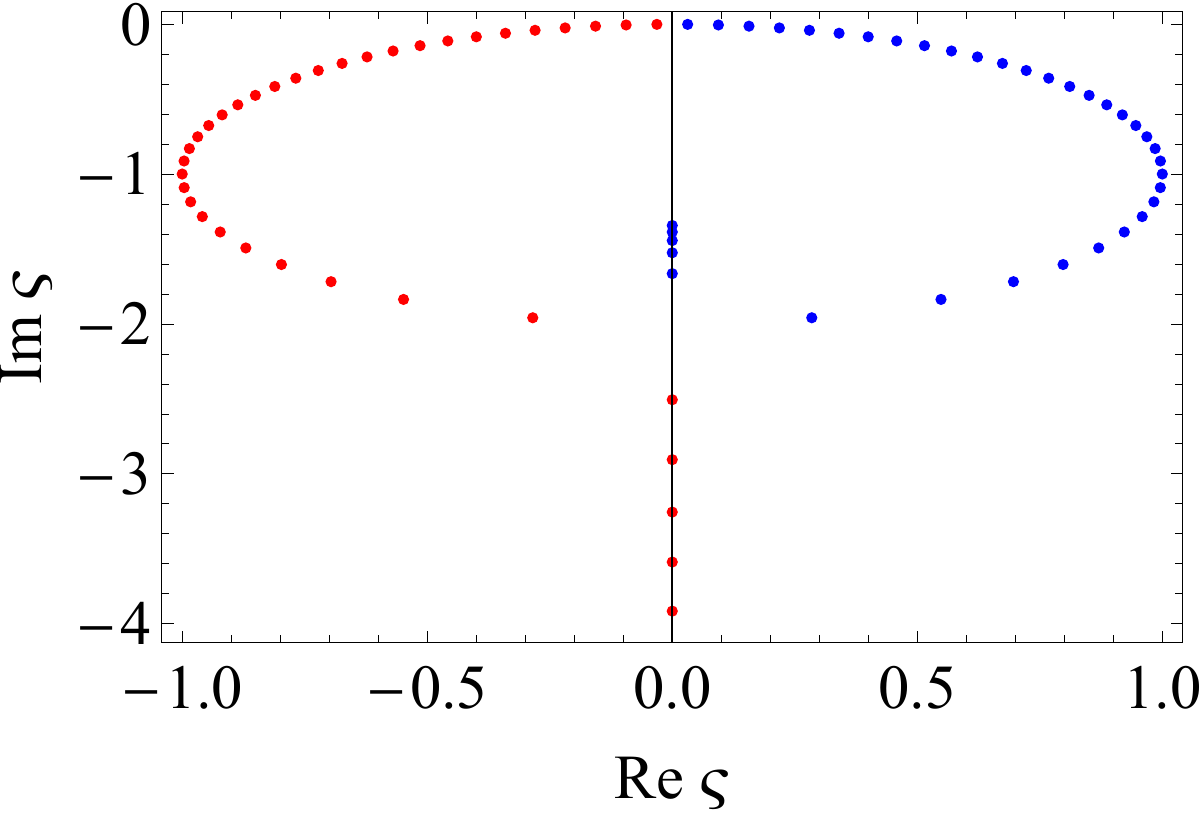}
\caption{Poles of the longitudinal stress response functions  in the complex frequency ($\varsigma$) plane for $\hn=50$ and $\gamman=2$. The poles are shown up to the order $|n|\leq 36$, of which the last four poles fall over the imaginary axis (become purely diffusive). The rest of the poles (shown in two left and right branches in red and blue colors, respectively) represent partially attenuated acoustic modes propagating across the  film. 
} 
\label{fig:poles_sn}
\end{center}
\end{figure}

For compressional modes, the roots of Eq. \eqref{roots_eq_app} can have both finite real and finite imaginary parts. The approximate loci of the roots  can be obtained  for $\Theta\gg 1$ and $\Theta\ll 1$ using limiting arguments analogous to those discussed above for the shear modes. Using the definitions in Eqs.  \eqref{eq:ellRdef}, \eqref{eq:ellIdef} and \eqref{eq:defs_for_roots_app}, we arrive at $\varsigma^2 + \icomplex \varsigma q_n^2 - q_n^2\simeq 0$,
solving which gives the roots $\varsigma_{\!_\perp}^{(n)}$ as 
\begin{equation}
\label{eq:sol_roots_norm_app2}
\varsigma_{\!_\perp}^{(n)}\simeq \pm\frac{|q_n|}{2} \sqrt{4- q_n^2 }-\icomplex\frac{q_n^2}{2},\\
\end{equation}
where we have  
\begin{equation}
\label{eq:qn_norm_app2}
 q_n=\frac{\pi}{\hn}\times 
\left\{\begin{array}{l l}
n+1/2&\,\,\, \hn/\gamman\gg 1,\\
n\,\, (n\neq 0)\,\,&\,\,\, \hn/\gamman\ll 1.
\end{array}\right.
\end{equation}
These poles exhibit a finite real part (representing sound propagation with a finite phase velocity) and a finite imaginary part (representing a finite lifetime due to sound absorption) for $|q_n|< 2$, or, equivalently,  $|n+1/2|<2\hn/\pi$, when $\hn/\gamman$ is sufficiently large.  These poles exhibit two distinct, mirror symmetric, left and right  branches with negative and positive  real parts, shown in Fig. \ref{fig:poles_sn} by red and blue symbols, respectively. The poles with larger order in $n$ become purely imaginary and the corresponding compressional modes become diffusive. The diffusive poles stemming from the right branch densely accumulate in a short interval just below $-\icomplex$, and gradually approach the latter as a limiting value, when $|n|\rightarrow \infty$; while, those stemming from the left branch  spread with increasingly large separations over the imaginary axis, tending to $-\icomplex\infty$. The real part of the poles never exceeds one in magnitude on either branch, irrespective of the other parameter values; thus, all of the poles fall within the frequency integration domain $|\omegan|\leq \omegan^\infty= 1$.

\begin{figure}[t!]
\begin{center}  		
\includegraphics[width=5.65cm]{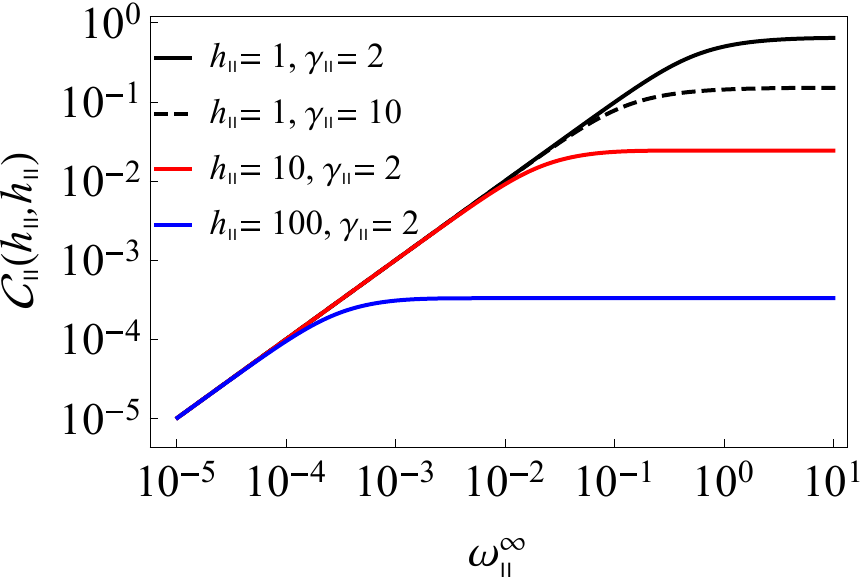} (a)
\\
\vspace{3mm}
\includegraphics[width=5.65cm]{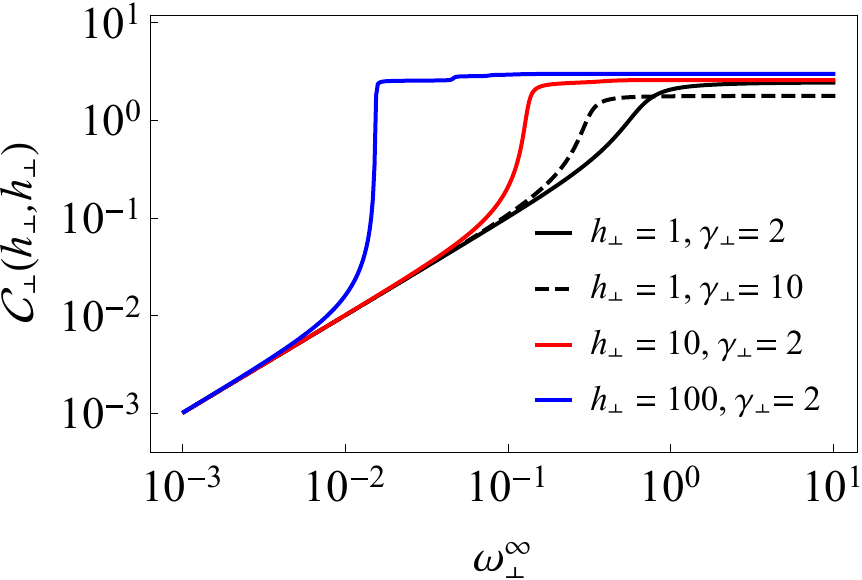} (b)
\caption{Log-log plots of equal-time self-correlators on the fixed plate, $\cshear(\hs,\hs)$ and $\cnorm(\hn,\hn)$, in panels (a) and (b) respectively, as functions of the corresponding dimensionless frequency cutoffs for different values of the dimensionless system parameters as indicated on the graphs.} 
\label{fig:cutoff_n}
\end{center}
\end{figure}

\section{Cutoff and other parameter values}
\label{app:cutoff}

Although the frequency integrals appearing in the analytical results are originally defined over the whole real axis, it is clear that the hydrodynamic description breaks down for very large frequencies, or at very small length-scales below a microscopic cutoff $a$. In our numerical analysis, we impose an upper frequency cutoff $\omega^\infty$ and take the integrals over the range  $|\omega| \leq \omega^\infty$; for consistency, we also limit the range of film thicknesses to $h\geq a$. Nevertheless, it is important to note that the frequency integrals in our analysis of the stress correlators, Eqs. \eqref{eq:shear_equal_time_redef} and \eqref{eq:norm_equal_time_redef}, are found to converge sufficiently rapidly, giving finite results as the cutoff is taken to infinity. The introduction of a cutoff is  important as it enables us to ensure that our results are primarily determined by the frequencies that fall within the hydrodynamic regime. 

For the most part, we set the microscopic length-scale cutoff  $a=\nusn/c_0$ with the corresponding frequency cutoff  $\omega^\infty = c_0^2/\nusn$ for the shear/compression modes. In dimensionless units, we have $\asn = 1$ and $\omegasn^\infty=1$ (Section \ref{subsec:dimless}). 
For the parameter values relevant to water, e.g., $\eta\simeq 8.9\times10^{-4}\,{\textrm{Pa}}\cdot{\textrm{s}}$, $\rho_0\simeq 997\,{\textrm{Kg}}\cdot{\textrm{m}}^{-3}$, $c_0\simeq 1496.7\,{\textrm{m}}\cdot{\textrm{s}}^{-1}$ at $25^{\circ}{\textrm{C}}$ \cite{CRC-Handbook} 
and assuming $\zeta\simeq 3\eta$ \cite{Dukhin2009}, the length-scale and frequency cutoffs for the shear (compression) modes are found approximately $a\simeq 0.6\,{\textrm{nm}}$ ($2.6\,{\textrm{nm}}$) and $\omega^\infty\simeq 2.5\times10^{12}\,{\textrm{s}}^{-1}$ ($0.6\times10^{12}\,{\textrm{s}}^{-1}$), respectively. 
These values agree with recent studies indicating validity of the continuum hydrodynamic description in water down to scales of around one nanometer \cite{Bocquet2010}. 
In other words, in the present context, taking the dimensionless frequency cutoffs larger than $\omegasn^\infty\sim 1$ may not be justified.  

We show the exemplary cases of equal-time self-correlators on the fixed plate, $\cshear(\hs,\hs)$ and $\cnorm(\hn,\hn)$, as  functions of the corresponding dimensionless frequency cutoffs in Fig. \ref{fig:cutoff_n} for selected  values of the dimensionless system parameters as indicated on the graphs. As seen, the plotted quantities saturate to a limiting cutoff-independent plateau when the corresponding cutoff is large enough. For the slowest rate of convergence to the plateau levels, which is obtained for the lowest admissible value of the inter-plate separation $\hs = \hn =1$ (or  $h=a$), we find that the saturations in $\cshear(\hs,\hs)$ and $\cnorm(\hn,\hn)$ occur roughly around $\omegas^\infty=\omegan^\infty=1$, when $\gammas=\gamman=2$. The results for larger values of $\hs$ and $\hn$ and/or larger values of $\gammas$ and $\gamman$ show faster convergence trends to their cutoff-independent values. The saturations occur roughly at $\omegas^\infty\sim \hs^{-2}$ ($\omegan^\infty\sim \hn^{-1}$) for the transverse (longitudinal) correlators.
These observations confirm that our numerical choices for the length/frequency cutoffs are consistent with those expected experimentally \cite{Bocquet2010} and, yet, they are large enough to give the desired saturation (plateau) values for the numerical outcomes. 

It is also to be noted that for longitudinal correlators, the saturation is rather abrupt, occurring when the cutoff $\omegan^\infty$ exceeds the locus of the first peak along the real-frequency axis (panel a in Fig. \ref{fig:Cn_freq}). In other words, the longitudinal frequency integrals quickly saturate to their cutoff-independent values as soon as the integration domain, $|\omegan| \leq \omegan^\infty$, is wide enough to include the first peak (first excited acoustic mode), as this peak gives the major dominant contribution to the frequency integrals.



\begin{thebibliography}{99}
\bibitem{Note_small_parameter}
To avoid any possible inconsistencies, we assume that the supremum norms of the surface velocity $|{\mathbf u}|$, external surface forcing $|{\mathbf f}|$ and film thickness variations $|\delta h|$ are all sufficiently small and of the same order. The film thickness variations ($|\delta h|\ll h$) can be shown to contribute only to the subleading terms in our calculations.
\bibitem{Note_T}
Strictly speaking, ignoring local temperature variations and heat transfer processes is equivalent to setting the thermal conductivity coefficient, the isobaric thermal expansion coefficient and specific heat at constant volume equal to zero \cite{LL,Boon-Yip,Berne-Pecora}. 
It turns out, however, that only setting the isobaric thermal expansion coefficient equal to zero is enough to ensure that the local temperature does not vary up to the leading order in fluctuations.
\bibitem{Note_typo}
We note the typographic error in Ref. \cite{Monahan}, referring to $c_0$ as adiabatic speed of sound.
\end{thebibliography}
\end{document}